\documentclass{article}
\usepackage{color,graphicx}
\usepackage[latin1]{inputenc}
\usepackage[authoryear]{natbib}
\usepackage{amsmath,amsfonts,amssymb,amsthm}
\usepackage{fontenc,url,path}
\usepackage{lscape,epsfig}
\usepackage{hyperref}
\hypersetup{colorlinks,
            urlcolor=blue,
            citecolor=blue}
\usepackage{booktabs}
\usepackage{natbib}
\usepackage{dsfont} 
\newcommand{\bm}[1]{\mbox{\boldmath $#1$}}

\newcommand{\keywordname}{Key Words}
\newcommand{\ackname}{Acknowledgements}

\newcommand{\keywords}[1]{%
    \begin{quote}
    \small\textbf{\keywordname: }{#1}
    \end{quote}
}
   {}{}

\definecolor{dgreen}{rgb}{0.,0.6,0.}
\newcommand{\bjb}{\color{black}} 
\newcommand{\skadd}[1]{\textcolor{black}{#1}} 
\newcommand{\pmr}[1]{\textcolor{black}{#1}} 

\RequirePackage[normalem]{ulem} 
\RequirePackage{color}\definecolor{RED}{rgb}{1,0,0}\definecolor{BLUE}{rgb}{0,0,1} 

\begin{document}

\graphicspath{{Plots/}}

\title{\Large\bf Model selection and parameter inference in phylogenetics
	using Nested Sampling}

\author{Patricio Maturana R.${}^1$\thanks{Corresponding author. Address for					correspondence: Department of Statistics, University of Auckland, Private Bag 92019, Auckland 1142, New Zealand. Email address: \href{mailto:p.russel@auckland.ac.nz}{p.russel@auckland.ac.nz} (Patricio Maturana R.)}
	\quad Brendon J. Brewer${}^1$ \quad Steffen Klaere${}^{1,2}$ \quad Remco Bouckaert${}^{3,4}$ \\[1.2ex]
	{\footnotesize ${}^1$Department of Statistics, University of Auckland, Auckland, New Zealand} \\
	{\footnotesize ${}^2$School of Biological Sciences, University of Auckland, Auckland, New Zealand}\\
	{\footnotesize ${}^3$ Center of Computational Evolution, University of Auckland, Auckland, New Zealand}\\
	{\footnotesize ${}^4$Max Planck Institute for the Science of Human History, Jena, Germany}
 }

\date{}
\maketitle




\begin{abstract}

Bayesian inference methods rely on numerical algorithms for both model selection and parameter inference.  In general, these algorithms require a high computational effort to yield reliable estimates.  One of the major challenges in phylogenetics is the estimation of the marginal likelihood.  This quantity is commonly used for comparing different evolutionary models, but its calculation, even for simple models, incurs high computational cost. Another interesting challenge relates to the estimation of the posterior distribution. Often, long Markov chains are required to get sufficient samples to carry out parameter inference, especially for tree distributions.  In general, these problems are addressed separately by using different procedures.  Nested sampling (NS) is a Bayesian computation algorithm which provides the means to estimate marginal likelihoods together with their uncertainties, and to sample from the posterior distribution at no extra cost.  The methods currently used in phylogenetics for marginal likelihood estimation lack in practicality due to their dependence  on many tuning parameters and the inability of most implementations to provide a direct way to calculate the uncertainties associated with the estimates.  To address these issues, we introduce NS to phylogenetics.  Its performance is assessed under different scenarios and compared  to established methods.  We conclude that NS is a competitive and attractive algorithm for phylogenetic inference. An implementation is available as a package for BEAST 2  under the LGPL licence, accessible at \url{https://github.com/BEAST2-Dev/nested-sampling}.

\keywords{model selection, parameter inference, nested sampling, marginal likelihood}

\end{abstract}




Bayesian methods provide a comprehensive framework in which to explore parameter space, uncertainty and model-to-data fitness. The concept was introduced to phylogenetics in the 1990s \citep{Rannala:Yang:1996, Yang01071997, Larget:1999}, and has gained popularity because of its flexibility when dealing with complex models and large data sets, in contrast with maximum likelihood estimation.  The increase in computational power led to the rise of Bayesian methods, as Markov Chain Monte Carlo (MCMC) approaches became feasible. Its popularity was further increased by state-of-the-art implementations of the models in programs like MrBayes \citep{MrBayes:2001}, BEAST
\citep{Drummond:2007jk,BEAST2}, and PhyloBayes \citep{PhyloBayes:2009}.

As in other fields, model selection plays an integral part in phylogenetic inference. A wide variety of different criteria are available for this task, with some of the most popular being the Likelihood Ratio Test (LRT), Akaike Information Criterion (AIC), Bayesian Information Criterion (BIC), and Bayes Factors (BF), {\bjb which are ratios of marginal likelihoods}.  The latter is of particular interest as it provides many advantages over the other methods: i) It is a direct consequence of probability theory used as a theory of reasoning; ii) it allows comparison of nested and non-nested models; iii) it is not based on a point estimate in parameter space since it averages over parameter space; and iv) it embodies Occam's razor by involving the prior distribution in the model selection process. {\bjb Marginal likelihoods penalise the inclusion of a new parameter when its value is unknown and some of the possible values do not fit the data well}.  However, {\bjb the marginal likelihood} is a difficult integral that depends on the complexity of the model. Finding ways to efficiently estimate this integral is one of the major challenges of the field.


{\bjb A simple} Monte Carlo method for estimating the marginal likelihood is the harmonic mean \citep{Newton:Raftery:1994}. Despite its popularity, it is well-known to overestimate the real value, and its variance is infinite in most situations, resulting in unreliable estimates.  Among the most accurate methods currently used in phylogenetics are path sampling \citep{Lartillot:Philippe:2006} and stepping-stone sampling \citep{Xie:Lewis:Fan:Kuo:Chen:2011} {\bjb which are much more accurate but have} high computational cost.  The latter has gained popularity in recent years due to its implementation in different phylogenetic software packages, such as  MrBayes \citep{MrBayes:2001} and BEAST \citep{Drummond:2007jk,BEAST2}.  However, these  methods have a relatively large number of {\bjb tuning} parameters that need to be set prior to analysis, and there is no rigorous method of determining the values appropriate for the accurate estimation of the marginal likelihood. Also, these methods have problems dealing with some likelihood shapes \citep{Skilling:2006}.

To efficiently deal with the above issues a generalised version of stepping-stone sampling (GSS) has been proposed \citep{gss:2011}. This generalisation also {\bjb allows us} to regard the phylogeny as an unknown parameter \citep{Holder:2014, Baele:2016} {\bjb incorporating} the uncertainty in {\bjb the} tree topology in model selection.

A more general technique for the estimation of the marginal likelihood is nested sampling \citep[NS;][]{Skilling:2006}. This method requires less tuning and can deal with partly convex likelihood functions (henceforth, in the NS sense, explained below).  Its main feature is the reduction of the multidimensional integral over parameter space to a one-dimensional integral {\bjb of the likelihood as a function of the enclosed prior probability}. This technique, and several variants \citep[e.g.][]{MultiNest:2009, Brewer:2011, Handley:2015} have been successfully applied to fields like astronomy \citep{Mukherjee:Parkinson:Liddle:2006, Brewer:Donovan:2015} {\bjb and} systems biology \citep{Aitken:Akman:2013, Pullen:Morris:2014} and have shown great promise in parameter inference and model selection.

In this paper, we assess the merits of nested sampling in phylogenetic inference, and compare it to established methods.  \pmr{Firstly, the method is assessed in a small phylogenetic test case}.  Secondly, a reasonably big dataset, which contains several partitions and many parameters, is analysed in order to show the consistent marginal likelihood estimates and parameter inferences yielded by NS.  Finally, two datasets, which have become standard phylogenetic datasets for the analysis of MCMC method performance, are analysed for marginal likelihood estimation and parameter inference.


\section{Bayesian inference}

Let $\bm{\theta}$ be the vector of parameters, $\bm{X}$ the data, and $M$ the model (assumed throughout). Bayes' theorem is given by
\begin{align}
\label{eq:post_theta}
f(\bm\theta | \bm X, M) = \frac{L(\bm X |\bm\theta, M) \pi(\bm\theta|
	M)}{f(\bm X | M)}.
\end{align}
The prior distribution $\pi(\bm\theta| M)$ represents our previous knowledge of the parameters which is updated after taking into account the {\bjb data; the updated knowledge is} reflected in the posterior probability distribution $f(\bm\theta | \bm X, M)$.  The likelihood function $L(\bm X |\bm\theta, M)$ represents the probability of the data given the parameters and the model.  The marginal likelihood $f(\bm X | M)$ is the probability of the data under the model and plays a key role in model selection. Indeed, this quantity is used to select among models. Because of this, it is also called the \textsl{evidence} \citep*{MacKay:2002}. To understand its role, note that the posterior distribution for the model $M_j$ is given by
\begin{align*}
f(M_j | \bm X) = \frac{f(\bm X | M_j) f(M_j)}{f(\bm X)}, \quad j=0,1,
\end{align*}
where $f(\bm X | M_j)$ is the marginal likelihood as shown in \eqref{eq:post_theta}, $f(M_j)$ is the prior probability for the model, and $f(\bm X)$ is the probability of the data. The marginal likelihood will also be denoted by ``$\mathcal{Z}$" henceforth.  The Bayesian comparison of two models $M_0$ and $M_1$ can be carried out by comparing their posterior probabilities.  This comparison is {\bjb often} through the ratio of their probabilities, which represents the plausibility of one model over another and is defined as follows:
\begin{align*}
\dfrac{f(M_0 | \bm X)}{f(M_1 | \bm X)} &= \frac{f(\bm X | M_0)}{f(\bm X | M_1)}
\dfrac{f(M_0)}{f(M_1)},\\
\text{posterior odds} &= \text{Bayes factor} \times \text{prior odds.}
\end{align*}
The ratio of marginal likelihoods, the first ratio on the right side, is called the Bayes factor \citep{Kass:Raftery:1995}. If we have  no preference for any model, i.e., each model is assigned the same prior probability,
the priors cancel each other out and the posterior odds is only given by ratio of the marginal likelihoods.

Although the marginal likelihood is generally ignored in parameter inference, it plays a key role in model selection: it is a measure of the goodness of fit.  Indeed, it is the probability of the data given the model, i.e., it is by definition a measure of model fit.  The marginal likelihood acts as the normalisation constant in the posterior distribution making it a probability density function.  Thus, this quantity is a multidimensional integral of the prior distribution times the likelihood function over the parameter space.   MCMC methods used for parameter estimation within a model use only ratios of posterior densities, and are therefore unable to measure its normalisation in general.

Unlike maximum likelihood, which represents the model fit at a single point, this quantity stands for an average of how well the model fits the data.  By being an average of the likelihood function with respect to the prior, the model with the greatest evidence might be different from the model with the highest likelihood because the prior could down-weight some regions of parameter space.  Also, the marginal likelihood is sensitive to the size of the region over which the likelihood is high.  As a result, both methods could favour different models.  Despite its important role in model selection, the marginal likelihood is usually analytically intractable and has to be approximated by numerical methods.

\subsection{Estimation of marginal likelihoods}

Typically, phylogenetic models involve a high level of complexity, making it difficult to calculate the marginal likelihood.
\citet{Suchard:Weiss:Sinsheimer:2001} proposed the Savage-Dickey ratio to
estimate Bayes factors for nested models \citep{Verdinelli:1995}.  \citet{Huelsenbeck:2004} used reversible jump Markov chain Monte Carlo including all possible time-reversible models.  Nevertheless, these methods are restricted to a particular group of models. Other alternatives have been proposed to allow a more general comparison of models. Among them, the harmonic mean (HM) is the most popular to estimate the marginal likelihood \citep{Newton:Raftery:1994}, an importance-sampling approach.  Its popularity is due to its simplicity, it only requires samples from the posterior distribution.  However, the HM estimator often has infinite variance, overestimates the true value of the marginal likelihood, and does not work in high dimensions, the usual case in phylogenetics \citep{Newton:Raftery:1994,Lartillot:Philippe:2006, Xie:Lewis:Fan:Kuo:Chen:2011, ssbf:2013, Baele:2012, Baele:2016}.

Far more accurate than the HM method is path sampling (PS), also known as thermodynamic integration, proposed in phylogenetics by \cite{Lartillot:Philippe:2006}.  The method requires several Markov chains from transition {\bjb distributions} which {\bjb form} a path between the prior and the posterior distribution.  These transition functions are defined by the ``power posterior"
\begin{align*}
p_{\beta} = \frac{L(\bm X |\bm\theta, M)^{\beta} \pi(\bm\theta| M)}{\mathcal{Z}_{\beta}},
\quad \text{for} \quad 0 \leq \beta \leq 1,
\end{align*}
where $\mathcal{Z}_{\beta}$ is the normalising constant of the unnormalised power posterior density $L(\bm X |\bm\theta, M)^{\beta} \pi(\bm\theta| M)$.  Similarly, a path between the posterior of two models could be defined to estimate the Bayes factor directly.  Note that for $\beta = 0 $ the power posterior is equivalent to the prior distribution and for $\beta = 1$ is equivalent to the posterior distribution.  In the latter case, $\mathcal{Z}_1=\mathcal{Z}$ is the marginal likelihood.  PS relies on the identity
\begin{align*}
\log \mathcal{Z}=\int_{0}^{1}\mathds{E}_{p_{\beta}}\big[\log L(\bm X|\bm\theta, M)\big ] \text{d}\beta,
\end{align*}
where the expected value is with respect to the power posterior distribution $p_{\beta}$.  PS uses a series of $\beta$ values which define the transition distributions.  For each value, a Markov chain is required to estimate the expected values and consequently the integral {\bjb over $\beta$}. Clearly, the increase in accuracy comes at the cost of increased computational complexity.

Another importance sampling approach is stepping-stone sampling (SS) proposed by \citet{Xie:Lewis:Fan:Kuo:Chen:2011}.  SS relies on transition {\bjb distributions} like PS in order to define an equivalence between the marginal likelihood and the telescope product of ratios of normalising constants given by
\begin{align*}
\mathcal{Z} = \prod_{k=1}^{K} \dfrac{\mathcal{Z}_{\beta_{k}}}{\mathcal{Z}_{\beta_{k-1}}},
\end{align*}
where $\beta_0 = 0 < \beta_1 < \dots < \beta_{K-1} < \beta_K = 1$.  Each ratio $\mathcal{Z}_{\beta_{k}}/\mathcal{Z}_{\beta_{k-1}}$ is estimated by importance sampling. The performance of this method is similar to PS.  However, SS requires slightly less computational effort: it does not need posterior samples and in general requires a smaller number of transition distributions to reduce its discretisation bias than PS \citep{Xie:Lewis:Fan:Kuo:Chen:2011}.  SS also allows us to estimate the Bayes factor directly defining a path between the posterior of both models \citep{ssbf:2013}. The extended version of SS, generalised stepping-stone sampling \citep[GSS;][]{gss:2011},  uses a reference distribution to shorten the distance between the prior and posterior distribution.  This strategy can potentially lead to a more efficient estimation process.  Like in its predecessor, the geometric path is often used to connect these densities, which is defined by
\begin{align*}
p_{\beta} = \frac{\big (L(\bm X |\bm\theta, M) \pi(\bm\theta| M)\big)^{\beta}
	\pi_{0}(\bm\theta| M)^{1-\beta} }{\mathcal{Z}_{\beta}}, \quad \text{for} \quad 0 \leq \beta \leq 1,
\end{align*}
where $\pi_0$ is the reference distribution.  When $\beta = 0$ and $\beta = 1$, the power posterior is equivalent to the reference and posterior distribution, respectively.  GSS is more accurate than the original version when the reference distribution approximates {\bjb the posterior distribution reasonably well}.   This method has also been extended, allowing the tree topology to be variable \citep{Holder:2014, Baele:2016}.  This allows {\bjb the user} to accommodate phylogenetic uncertainty in model selection.  GSS has the potential of leading to remarkable improvements in comparison to the original SS and PS, that is, less tuning parameters, lower variance, avoidance of numerical instabilities, reduction in the computational time, and it is more accurate in case of very diffusive
priors, in which cases PS/SS overestimate the true marginal likelihood \citep{Baele:2016}.

Although PS and SS yield accurate estimates of the marginal likelihood, they require several specifications depending on the problem.  Firstly, an annealing schedule (a number of $\beta$-values) is required.  A common practice is to try with different numbers until the estimate is stable.  This commonly used procedure is described in \citet{drummond2015bayesian} as follows: ``\textit{run the path sampling analysis with a low number of steps (say 10) first, then increase the number of steps (with say increments of 10, or doubling the number of steps) and see whether the marginal likelihood estimates remain unchanged}".  This could be impractical in some situations, for instance, when flat priors are used, which would increase the number of steps, or for big datasets.  Actually, flat priors are most often incorrectly used and constitute improper priors, challenging MCMC sampling \citep{Baele:2013}.  Secondly, the path described by the $\beta$-values has to be defined. \citet{Lartillot:Philippe:2006} proposed to spread the $\beta$-values regularly spaced between 0 and 1.  But since often most of the variability of the expected values is concentrated for $\beta$ near 0, some authors have proposed to concentrate the computational effort in that place.  For example \citet{Lepage:2007} used a sigmoidal function to estimate the Bayes factor using PS; \citet{Friel:2008} proposed $\beta_k = x_{k}^{4}$ in PS, where $x$-values are equally spaced between 0 and 1; and  \citet{Xie:Lewis:Fan:Kuo:Chen:2011} advocated spreading the values according to evenly spaced quantiles of a Beta$(\alpha,1)$, with $\alpha = 0.3$.  Finally, these methods require a number of samples from the power posterior for each $\beta$-value.  The main problem is that optimal settings vary from case to case.  The popularity of SS is due to its implementation in popular software such as MrBayes \citep{MrBayes:2001} or BEAST \citep{Drummond:2012}.  Either, the settings have to be defined by the user, or some predetermined tuning parameters can be chosen, but those may be unsuitable for the problem at hand.

In this context, GSS with an appropriate reference distribution requires potentially fewer tuning parameters.  Firstly, it requires an annealing/melting scheme  (a number of $\beta$-values).  The estimation can start from either the prior or posterior distribution.  Unlike SS or PS, the $\beta$-values do not necessarily need to follow any particular distribution to effectively control the uncertainty of the estimate, because of the similarity of the reference and posterior distributions \citep{gss:2011}.  Thus, the values can be equally spaced between 0 and 1.  Also, GSS does not need as many transitional distributions as its original version and it is more robust to prior specifications, i.e., the prior does not have a huge effect on the method performance. Finally, the method requires a number of samples from each transitional distribution.

PS and SS have usually been presented as methods of general applicability \citep{Xie:Lewis:Fan:Kuo:Chen:2011,ssbf:2013,Arima:2014,Baele:2014}. However, these methods only work when the shape of the likelihood, as a function of the cumulative prior probabilities (see Figure~\ref{Figure1}), is concave.  Partly convex likelihood functions might make them require impractical computational effort or make them fail outright \citep{Skilling:2006}.  The transition distributions are unable to mix between different phases of the likelihood function, resulting in a poor estimate.  \pmr{GSS is an efficient alternative but works well if and only if an appropriate reference distribution is used, otherwise it can fail dramatically (see \hyperref[ex:stat_example]{Appendix} for an example of this situation).}  A more general method is nested sampling \citep{Skilling:2006}, an algorithm that measures the relationship between likelihood values and the prior distribution, and uses this to compute the marginal likelihood.  This characteristic allows it, among other things, to cope with partly convex likelihood functions.  More importantly, unlike PS, SS and GSS, NS requires less problem-specific tuning.


\section{Nested Sampling}

Here we explain nested sampling in more detail than the original paper \citep{Skilling:2006} and give details on application to phylogenetic inference.
The marginal likelihood or evidence, in simplified notation, is given by \begin{align}
\label{eq:def.evidence}
\mathcal{Z}= \int_{\Theta} \pi(\bm\theta) L(\bm{\theta}) \text{d}\bm{\theta},
\end{align}
where $\bm{\theta} \in \Theta$ is the parameter vector, $L(\bm{\theta})$ is the likelihood function and $\pi(\bm\theta)$ is the
prior distribution. All the conditionals have been omitted, that is, $L(\bm{\theta})$ is written as the likelihood function instead of $L(\bm{\theta}|\bm{X}, M)$ and $\pi(\bm\theta)$ is written as the prior instead of $\pi(\bm\theta|M)$.

This definition applies for a continuous parameter space $\Theta$ where the phylogeny is assumed to be fixed.  When the tree topology is unknown, the parameter space is additionally composed by a discrete part.  In this case, the marginal likelihood incorporates the {\bjb sum} over the tree parameter space and is known as \textit{total marginal likelihood}.  Strictly speaking, its definition is given by
\begin{align*}
\mathcal{Z}= \sum_{\tau \in \mathcal{T}} \int_{\mathcal{V}_{\tau}} \int_{\Theta} L(\bm X |\bm\theta, \nu_{\tau}, \tau, M) \pi(\bm\theta, \nu_{\tau}, \tau| M) \text{d}\bm{\theta} \text{d} \nu_{\tau},
\end{align*}
where $\bm{\theta} \in \Theta$ is the parameter vector composed by elements such as frequencies, gamma parameter and rates parameters, $\nu_{\tau} \in \mathcal{V}_{\tau} $ is the set of branch lengths of $\tau \in \mathcal{T}$ which is the tree topology, $\bm X$ is the molecular data and $M$ is the substitution model.  For simplicity, NS will be described for definition \eqref{eq:def.evidence}, but its generalisation to the case of variable tree topology is analogous.

\begin{figure}
	\centering
	\includegraphics[scale=0.50,clip=true,angle=0]{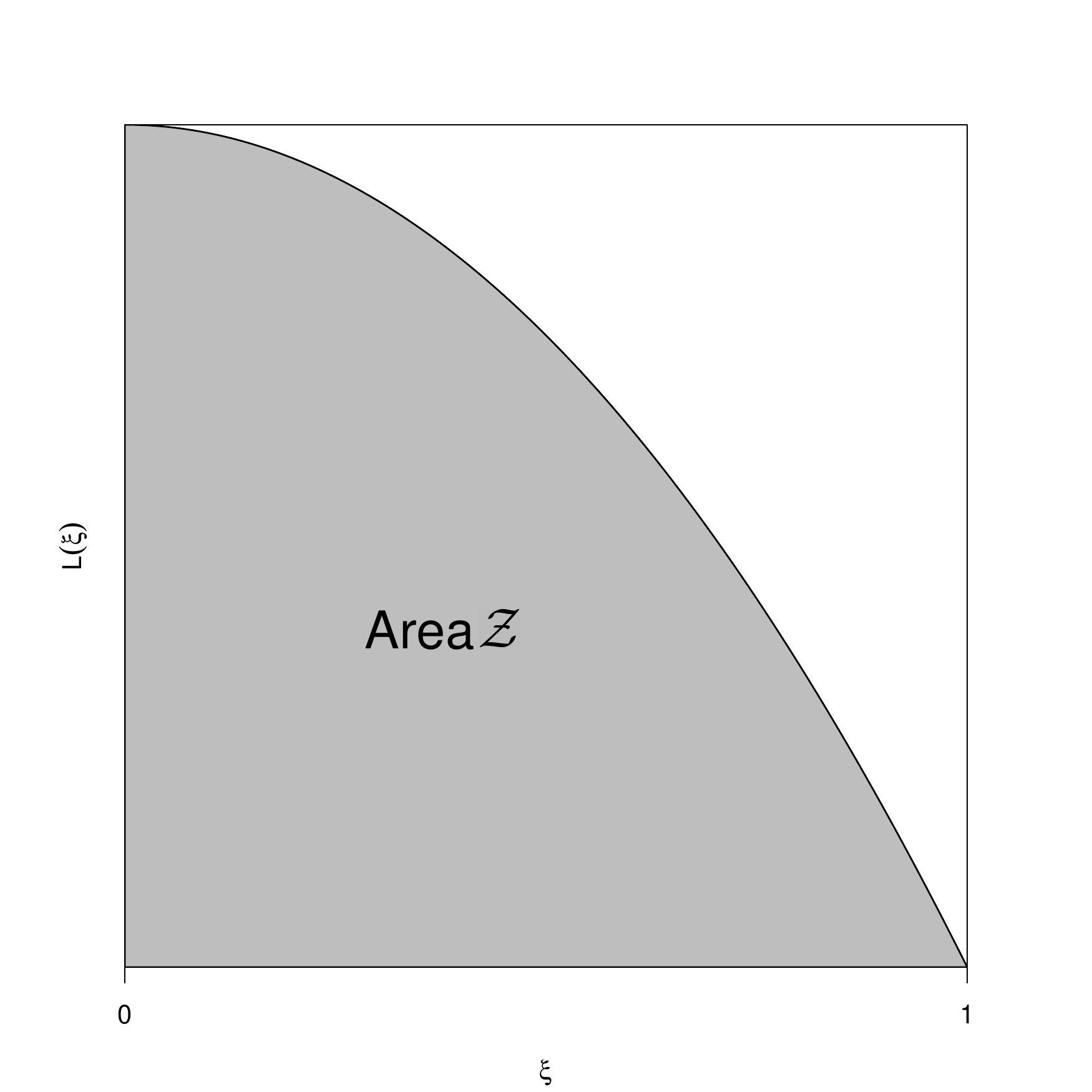}
	\caption{Association between the cumulative prior mass and the likelihood function.  Nested sampling estimates the gray area which is the marginal likelihood.  In general, a small area of the prior concentrates high likelihood values causing the area to be concentrated around $\xi \approx 0$. }
	\label{Figure1}
\end{figure}

To understand the key idea of NS, consider that for any positive random variable $Y$, its expected value can be written as
\begin{align*}
\mathds{E}[Y] = \int_{0}^{\infty} \big(1-F(Y)\big)\text{d}Y,
\end{align*}
{\bjb which} represents the area between the cumulative distribution function of $Y$, $F$, and 1.  Similarly, the likelihood function $L(\bm{\theta})$ can be seen as a positive random variable where $\bm{\theta}$ follows the prior distribution $\pi(\bm{\theta})$ and the evidence as the expected value of the likelihood function.  Nested sampling takes advantage of this property by transforming the multi-dimensional integral defined in \eqref{eq:def.evidence} into a one-dimensional integral as follows
\begin{align}
\label{eq:exp.like}
\mathds{E}_{\theta}\big[L(\bm{\theta})\big] \equiv \mathds{E}_{\lambda}\big[\lambda
\big] = \int_{0}^{\infty}\big(1-F(\lambda)\big)\text{d}\lambda,
\end{align}
where $\mathds{E}_{\theta}[\cdot]$ and $\mathds{E}_{\lambda}[\cdot]$ stand for the expectation with respect to the densities of $\theta$ and $\lambda$ respectively, $\bm\theta \sim \pi(\bm\theta)$, $\lambda = L(\bm\theta)$ and $F(\lambda)$ is the cumulative distribution function of the likelihood defined by
\begin{align*}
F(\lambda) = \idotsint \limits_{L (\scriptsize{\bm{\theta}})< \lambda}
\pi(\bm\theta) \text{d}\bm\theta.
\end{align*}
Let $\xi(\lambda)=1-F(\lambda)$ be the proportion of prior mass with likelihood greater than $\lambda$, and take its inverse, then the evidence given in \eqref{eq:exp.like} can be redefined as
\begin{align*}
\label{eq:int.z}
\mathcal{Z} = \int_{0}^{1} L(\xi) \text{d}\xi.
\end{align*}
This is the integral used by nested sampling, {\bjb and} is displayed in Figure~\ref{Figure1}. In general, this function concentrates its mass near zero because the posterior is located in a small area of the prior.  We use the {\bjb ``overloaded''} notation, where the same letter $L$ represents the likelihood function over different domains: $L(\bm{\theta})$ has the parameter vector $\bm{\theta}$ as argument, and $L(\xi)$ has the prior mass $\xi$ (scalar) as argument.  Note that $L(\xi)$ is a monotonically decreasing function which reaches its highest point at $\xi= 0$ and its lowest point at $\xi=1$ (see Figure~\ref{Figure1}).  $L(0.9) = 0.3$ means that $90\%$ of the draws $\bm{\theta}$ from the prior distribution will have likelihoods greater than 0.3. {\bjb If a set of points on the $L(\xi)$ curve can be obtained},
the integral can be approximated numerically by the basic standard quadrature method
\begin{equation}
\label{eq:sum.app}
\mathcal{Z} \approx \sum_{i=1}^{k} w_{i} L_{i},
\end{equation}
where $w_i= \xi_{i-1} - \xi_{i}$ (or $w_i = (w_{i-1}-w_{i+1})/2$ for the trapezoidal rule) and $L_i = L (\xi_i)$.  For a decreasing
sequence of $\xi$-values and an increasing sequence of $L$-values the evidence
can be estimated.  How to generate these sequences is described below.

\subsection{Sequence of $L$-values}

Nested sampling maintains a set of $N$ \textit{active points} $\bm\theta_1 , \dots , \bm \theta_N$ (with respective associated likelihood values $L(\bm\theta_1), \dots , L(\bm{\theta}_N)$) to generate the $i$th likelihood value required in \eqref{eq:sum.app}.  Initially they are drawn from the prior distribution, $\pi(\bm \theta)$.  From this set, the method requires selecting the point $\bm\theta_l$, where $l \in \{1, \dots, N\}$, with the lowest likelihood value.  
Then, the point $\bm\theta_l$ is discarded from the active points and replaced by a new point $\bm{\theta}$ sampled from the prior, {\bjb but} constrained to have a greater likelihood value than the {\bjb point being replaced, i.e.,} $L(\bm{\theta})> L(\bm{\theta}_l)$ (see Sampling section for more details).  This procedure shrinks the parameter space according to the likelihood restriction. The process is repeated until a given stopping rule is satisfied (more information on this will follow {\bjb later}).  Thus, a sequence of increasing likelihood values ($L_1, \dots, L_k$) and \textit{discarded points} ($\bm{\theta}_1,\dots, \bm{\theta}_k$) are generated. {\bjb The discarded points are the ones that contribute to the estimate of the marginal likelihood through their respective likelihoods.}


\subsection{Sequence of $\xi$-values}

The discarded points generate an increasing sequence of likelihoods, which are known precisely. An important insight of \citet{Skilling:2006} is that the corresponding $\xi$ values, while they cannot be measured precisely, can be estimated from the nature of the NS procedure.  The uncertainty of the NS estimate is mainly due to these approximations.

Nested sampling explores the prior distribution geometrically as follows
\begin{align}
\label{eq:geometric}
\xi_0 = 1, \:\: \xi_1 = t_1,\:\: \xi_2 = t_1 t_2, \quad \dots \quad , \:\: \xi_k =
\prod_{i=1}^{k} t_i,
\end{align}
where $t_i = \xi_{i}/\xi_{i-1} \in [0,1]$, for $i=1, \dots, k$.  This variable follows a $\text{Beta}(N,1)$ distribution.  This is because at the $i^{th}$ iteration, NS takes $N$ $x_i$ points which follows a $\text{Uniform}(0, \xi_{i-1})$, with $i = 1,\dots,N$.  These values are cumulative probabilities and consequently have a uniform distribution.  Their maximum value is $\xi_i$ which is related to the minimum likelihood value (note that $L(\xi)$ is a non-increasing function).  Since the distribution of $x_i / \xi_{i-1}$ is a $\text{Uniform}(0,1)$, their maximum value $\xi_i / \xi_{i-1}$ follows a $\text{Beta}(N,1)$ distribution.

{\bjb \citet{Skilling:2006} defined two schemes for estimating the $\xi$-values:} \textit{stochastic} and \textit{deterministic}.

\begin{itemize}
	\item \textit{Stochastic:} the $t_i$ values are generated randomly from the $\text{Beta}(N,1)$ distribution, for $i=1,\dots,k$.
	\item \textit{Deterministic:} the $t_i$ values are fixed by using their expectations as follows:
	\begin{itemize}
		\item Define its \textit{arithmetic mean}, $t_i=N/(N+1)$, {\bjb approximate} $\xi$-values would be given by
		$$\xi_i = \bigg(\dfrac{N}{N+1}\bigg)^i.$$
		\item Define its \textit{geometric mean}, $t_i = e^{-1/N}$, the estimated prior mass would be
		$$\xi_i=e^{-i/N}.$$
	\end{itemize}
\end{itemize}

Thus, a sequence of $\xi$ values can be generated and used in \eqref{eq:sum.app}.  The use of the geometric mean seems more reasonable given that the prior mass is defined geometrically by Equation \eqref{eq:geometric}. This scheme is considered for our examples, and is the one recommended by most authors.  However, the arithmetic mean allows nested sampling to be connected to rare event simulation \citep{Walter:2017}, and allows for an alternative version of NS with unbiased estimates of $\mathcal{Z}$.  On the other hand, the use of the stochastic approach has the potential of estimating more accurately the uncertainty by replicating the estimates for different $\xi$-sequences.


\subsection{Sampling}

The highest cost of nested sampling is in sampling from the restricted prior distribution (due to the condition that the likelihood needs to increase).  \cite{Skilling:2006} suggested to use a Metropolis-Hastings algorithm as usual, to explore the prior with the additional condition of rejecting the proposal points which do not fulfil the likelihood restriction.  As a starting value, a point from the sequence of active points can randomly be selected at each iteration of {\bjb NS, as all of them meet the likelihood condition by definition}.  Several other efficient methods have also been proposed \citep{Mukherjee:Parkinson:Liddle:2006, MultiNest:2009, Brewer:2011}.  We use \citeauthor{Skilling:2006}'s method to generate the restricted prior samples in our application.

Unlike the proposal mechanisms used in standard MCMC methods to sample the posterior, a static distribution, in NS such mechanisms have to deal with a variable target distribution over time.  In particular for tree proposals, this is a new scenario. Nested sampling compresses the prior at each iteration making it vary {\bjb at a constant rate}.  The proposals have to explore a wide area at the beginning which becomes constrained over time.  Tree proposal mechanisms should be able to adapt to this sampling characteristic.  Frequently, a uniform prior distribution is assigned over the tree parameter space which is quite huge even for few taxa.  Initially, bold moves would allow a good exploration using less steps than conservative ones.  However, the acceptance probability would decrease drastically over time due to the fact that the {\bjb target} distribution gets constrained.  On the other hand, conservative moves would require more steps to explore the prior distribution at the beginning, but later on, the acceptance probability would be higher than bold moves. Ideally, the proposal mechanism should take into account this dynamical behaviour of the target distribution over time.  \cite{Brewer:2016} stated that heavy-tailed proposals are as efficient as slice sampling \citep{Neal:2003}, at least in simple experiments.  
In this work we use the operators implemented in BEAST 2 \citep{BEAST2} and described in \cite{drummond2015bayesian}, but switched off auto-optimisation features, since the target distribution changes through time. 


\subsection{Information}

The idea of how much we have learned from the data is quantified through the notion of entropy.  The measure of information \citep{Sivia:Skilling:2006, Knuth:Skilling:2012} is given by the negative relative entropy
\begin{align*}
H = \int P(\bm{\theta}) \log
\bigg(\dfrac{P(\bm{\theta})}{\pi(\bm{\theta})}\bigg)\text{d}\bm\theta,
\end{align*}
where $P(\bm{\theta})$ and $\pi(\bm{\theta})$ are the posterior and prior distributions, respectively.  This quantity represents the amount of information in the posterior with respect to the prior, after acquiring data.  By definition, it can be seen as the expected value $H = \mathds{E}_P [\log (P(\bm{\theta})/\pi(\bm{\theta}))]$.  This can be approximated by
\begin{align*}
H \approx \sum_{i} \frac{w_{i} L_{i}}{\mathcal{Z}} \log \bigg(\frac{L_i}{\mathcal{Z}} \bigg)
\end{align*}
with $w_i = \xi_{i-1} - \xi_{i}$ \citep{Sivia:Skilling:2006}.  The following property of expected values is useful to understand the use of this concept. If $\text{G}_Y$ is the geometric mean of $Y$,  we have that
\begin{align}
\label{eq:geo.mean}
\log \text{G}_Y = \mathds{E}[\log Y] \Leftrightarrow \text{G}_Y =
e^{\mathds{E}[\log Y]}.
\end{align}
According to this property, $e^{-H}$ is a measure of central tendency or a typical value of $\pi(\bm{\theta})/P(\bm{\theta})$.  This value can be seen as the bulk of the posterior mass that occupies the prior. This idea helps to define a termination condition for nested sampling which will be described later. 

Note that a prior distribution which is consistent with the likelihood function, namely one that supports the same parameter values, has a lower information than one which likelihood function is in contradiction with the prior, i.e., their mass is concentrated in different places.  In other words, if the previous belief changes a lot after acquiring the data, more information has been gained from the data.

\subsection{Uncertainty}

The numerical uncertainty associated to the NS estimation of $\mathcal{Z}$ comes from two sources: i) approximating the prior volume ($w_i = \xi_{i-1} - \xi_{i}$), and ii) the error imposed by the integration rule.  However, the total uncertainty is usually dominated by the first.  Actually, the second is at most $\mathcal{O}(N^{-1})$ and $\mathcal{O}(N^{-2})$ for the simple standard quadrature and trapezoidal methods, respectively,  and thus negligible in comparison to the first source \citep{Skilling:2006}.

Thus, the uncertainty in $\log \widehat{\mathcal{Z}}$ depends directly on the uncertainty in $\sum_{i=1}^{k}\log \xi_i$.  Noting that $-\log \xi_i \sim \text{Exp}(N)$, and that consequently $-\sum_{i=1}^k \log \xi_i \sim \text{Gamma}(k, N)$, where $k$ is the number of iterations required by NS, we have that $\text{SD}\big[\sum \log\xi_i\big] = \sqrt{k}/N$.  \cite{Skilling:2006} argued that NS requires around $N \times H$ steps to reach the posterior, therefore its uncertainty can be approximated as
\begin{align}
\label{error}
\text{SD}\big[\log \mathcal{Z} \big] = \sqrt{\dfrac{H}{N}}.
\end{align}
The asymptotic variance of the nested sampling approximation grows linearly with the dimension of $\bm{\theta}$ and its distribution is asymptotically Gaussian \citep{Chopin:2010}.

Another way of calculating the uncertainty is by replicating the NS estimates for different $\xi$-sequences, i.e., using the stochastic approach, but keeping the same likelihood sequence.  Thus, a distribution of $\log  \widehat{\mathcal{Z}}$ can be inferred.  Note that this represents a marginal computational cost since most of it is spent by generating the likelihood sequence.  This strategy can be used similarly for parameter inference.


\subsection{Algorithm}

The algorithm iterates between the following steps:
\begin{enumerate}
	\item Sample $N$ points $\bm\theta_1, \ldots ,\bm\theta_N$ from the prior $\pi(\bm{\theta})$;
	\item Initialise $\mathcal{Z}=0$ and $\xi_0=1$;
	\item Repeat for $i=1, \ldots, k$;
	\begin{description}
		\item[i)] out of the $N$ live points, take the one with the lowest likelihood which we call $\bm{\theta}_{l}$ with corresponding likelihood $L_i=L(\bm\theta_{l})$,  where $l \in \{1, \dots, N\}$;
		\item[ii)] set $\xi_i=\exp(-i/N)$;
		\item[iii)] set $w_i= \xi_{i-1}-\xi_{i}$  (or $w_i = (\xi_{i-1}-\xi_{i+1})/2$ for the trapezoidal rule);
		\item[iv)] update $\mathcal{Z}= w_{i} L_{i} + \mathcal{Z}$; and
		\item[v)]  update the set of active points $\bm\theta_1, \ldots ,\bm\theta_N$ replacing $\bm\theta_l$ by drawing a new point $\bm\theta$ from the prior distribution restricted to $L(\bm\theta)> L(\bm\theta_{l})$.
	\end{description}
\end{enumerate}
Repeat the routine until a given stopping criterion is satisfied.  However, there is not a rigorous criterion that guarantees that we have found most of the bulk of $\mathcal{Z}$.  Nevertheless, some termination conditions have been proposed \citep{Skilling:2006}, which are discussed below.


\subsubsection{Termination}

The loop could continue until the potential maximum new contribution $L_i w_i$ represents a small fraction $\gamma$ of the accumulated evidence, that is
\begin{align*}
\max\big(L(\bm{\theta}_1), L(\bm{\theta}_2), \dots , L(\bm{\theta}_N)\big) w_i
< \gamma \mathcal{Z}.
\end{align*}
The algorithm can be stopped when the potential maximum new contribution is not significant.

Another criterion is based on the concept of information defined before. Typically, the likelihood values $L$ start dominating the prior mass $w$, so the contribution $L w$ increases at the beginning until the prior mass dominates this quantity.  After reaching a maximum, these values start to decrease.  The peak of this function is reached in the region of $\xi \approx e^{-H}$, when most of the posterior mass in the prior has been found.  Given that $\xi_i \approx e^{-i/N}$, a natural termination condition to estimate the log-evidence would be stopping the loop when $i/N$ significantly exceeds $H$, i.e., when the posterior mass has been explored completely.

There is no guarantee \textit{in general} that these termination conditions will work perfectly. $L$ might start increasing at a greater rate in the future, overwhelming the points that currently have high weights. In specific cases where the maximum likelihood value is known or can be roughly anticipated, it is possible to be confident that this won't happen.  In this work, we use the relative error as termination criterion.


\subsection{Posterior samples}
\label{post_samples}

NS yields posterior samples at no extra cost, if we assign appropriate weights to the discarded output points. In each iteration, NS has taken out a point from the active points generating a sequence of discarded points $\bm{\theta}_1, \bm{\theta}_2, \dots, \bm{\theta}_k$. These discarded points have contributed to estimate the marginal likelihood with their respective weights $w L$ which are proportional to the posterior distribution, in other words, prior {\bjb multiplied} by likelihood.  Thus, the sequence of discarded points can be sampled according to these weights in order to get a posterior sample.  The {\bjb effective sample size} (ESS) is related to the entropy of the posterior weights \citep{Skilling:2006} as
\begin{align*}
\text{ESS} = \exp \Bigg(- \sum_{i=1}^{k} p_i \log p_i \Bigg), \quad \text{where} \quad
p_i = \dfrac{w_i L_i}{\mathcal{Z} }.
\end{align*}




\section{Application}

The NS algorithm is assessed under different phylogenetic scenarios in order to show its performance for marginal likelihood estimation and parameter inference.  \pmr{Firstly,  NS is assessed in a 4 taxa case}.  Secondly, a reasonably big dataset \citep{Horn:2014} where the alignment has been split up into several partitions, each with their own site model, is used to carry out model selection via SS and NS.  In addition, it is used to assess the information provided by NS about the posterior distribution in comparison to a standard MCMC method.  Then, two datasets (Tetrapod and Chloridoideae) consisting of sequences of eukaryote species are analysed. These datasets form part of a group of standard datasets for evaluating MCMC methods \citep{Lakner:2008, Hohna:2012, Larget:2013, Whidden:2015} and possess interesting characteristics, challenging standard MCMC methods.  NS is performed using the deterministic approach to generate the $\xi$-sequence and the trapezoidal rule as method for the numerical integration.  We use the relative error as termination criterion with a error tolerance 1e-13.  All the analyses are performed in BEAST 2 \citep{BEAST2} and the plots produced in \textsf{R} \citep{R_project}.


\subsection{4 taxa example}

We want to verify that NS behaves well and is correctly implemented on a small example.
Since it is not possible to calculate marginal likelihoods analytically for phylogenetic datasets, even for simple models and simulated data, we study a 4~taxa case for which we can estimate its marginal likelihood with a high degree of confidence and thus evaluate NS.  The dataset, which is part of a bigger dataset available on Internet\footnote{\url{http://wiki.christophchamp.com/index.php?title=NEXUS_file_format}}, contains 4 taxa (Cow, Human, Mouse and Whale) with 705 nucleotides. An HKY and a strict clock model are assumed.  As prior for the tree, we consider a Yule model with a Normal($\mu=6.5, \sigma=0.1$) for its birth rate.  For the relative rate, a Lognormal($\mu=2.5, \sigma=0.1$).  These distributions are centred according to the posterior means and are tight in order to narrow the parameter space.

To have a reliable estimate of the true value of the marginal likelihood, we followed the analysis proposed in \cite{Baele:2016} and integrate the likelihood against the prior (ILP), popularly known as the arithmetic mean.  ILP consists in an average of likelihood values of a sample drawn from the prior.  This method produces an unbiased estimate.  For comparative purposes, we also consider the stepping-stone method.

For ILP we use a Markov chain of 2 billion samples, from which we take samples every 1,000 points.  For NS, we use 100 active points and 20,000 MCMC steps per iteration in order to to guarantee independence in the samples.  For SS, we use 100 steps.

The ILP estimate is -2349.97.  The average of 4 SS estimates is -2349.55.  The NS estimate is -2349.70 with a SD of 0.38. So, we conclude NS is in agreement with the ILP method, showing its consistency in a manageable small case.


\subsection{Euphorbia}

This dataset \citep{Horn:2014} contains 197 taxa with 6,328 nucleotides. Only chloroplast sites were used, divided into 15 partitions.  This represents a challenging case since it involves many parameters and consequently a huge parameter space.  We evaluate NS performance for model selection and compare it to SS results.  Also, we assess the ability of NS in sampling the posterior distribution and compare it to a standard MCMC method.

\subsubsection{Model selection}

Two clock models are compared: a \textit{strict clock model} and a \textit{relaxed clock model} \citep{Drummond:2006}.  These two clock models are compared through their marginal likelihoods.  First, we estimate these values by using SS with 400 steps and 750 samples per each transitional distribution.  These specifications have been tested and yield reliable estimates (results not shown).  The analysis is replicated using NS with a single active point and 30,000 MCMC steps per NS iteration, to generate the samples required.  \pmr{For the strict and relaxed clock models, NS took approximately 6.3 and 8.9 hours, respectively, on an Intel Core i5-7600 CPU @ 3.50GHz.}

The results are displayed in Table~\ref{Table1}.  SS estimates show the better fit of the relaxed clock model in comparison to the strict clock model; the former has a marginal likelihood substantially higher than the latter.  In terms of the Bayes factor, there is strong evidence in favour of the relaxed clock model \citep{Kass:Raftery:1995}.

NS is consistent with SS taking into account the uncertainty associated with the estimates.  Actually,  the 95\% confidence intervals contain the SS estimates in both cases.  The model selection can be made based on these intervals since they do not overlap.

This example shows the effectiveness of NS to carry out model selection.  In the hypothetical case in which the intervals would have overlapped, the analyses should have been redone but increasing the number of active points in order to decrease the uncertainty and consequently the width of the intervals.

\begin{table}
	\centering
	\scalebox{0.8}{
		\begin{tabular}{lcccccccc}
			\toprule
			Model &  \multicolumn{1}{c}{$N$} &  \multicolumn{1}{c}{$H$} &  \multicolumn{1}{c}{SD} &  \multicolumn{1}{c}{Iterations} &  \multicolumn{1}{c}{$\log \widehat{\mathcal{Z}}$} &  \multicolumn{1}{c}{lower} &  \multicolumn{1}{c}{upper} & \multicolumn{1}{c}{SS}\\
			\midrule
			Strict clock  & 1 & 1356.10 & 36.83	& 1,427 & -69611.60 & -69683.78 &-69539.42 & -69603.88 \\
			Relaxed clock & 1 & 1604.64 & 40.06 & 1,689 & -69100.21 & -69178.72 & -69021.70 & -69054.25 \\
			\bottomrule
		\end{tabular}
	}
	\caption{NS and SS marginal likelihood estimates for the Euphorbia dataset.  NS includes its 95\% confidence intervals, which contain the SS estimates and make evident its potential for model selection even under the simplest specifications (using a single active point).}
	\label{Table1}
\end{table}

\subsubsection{Parameter inference}

Recycling the NS run executed in the previous analysis, it is possible to carry out parameter inference.  For this, we recalculate the posterior weights for a new sequence of $\xi$-values (obtained from a Beta distribution), and generate a new posterior sample as described \pmr{above}.  The mean and the standard deviation are calculated from this sample.  This procedure was replicated 1,000 times registering these statistics. Note that this procedure is not computationally expensive since it does not require likelihood evaluations \pmr{(just a few seconds in \textsf{R})}. In addition, an MCMC analysis is performed and its statistics registered.  The chain length is 50,000,000 with a burn-in period of 10\% and thinning factor of 10,000.

\begin{figure}
	\centering
	\includegraphics[scale=0.45,clip=true,angle=0]{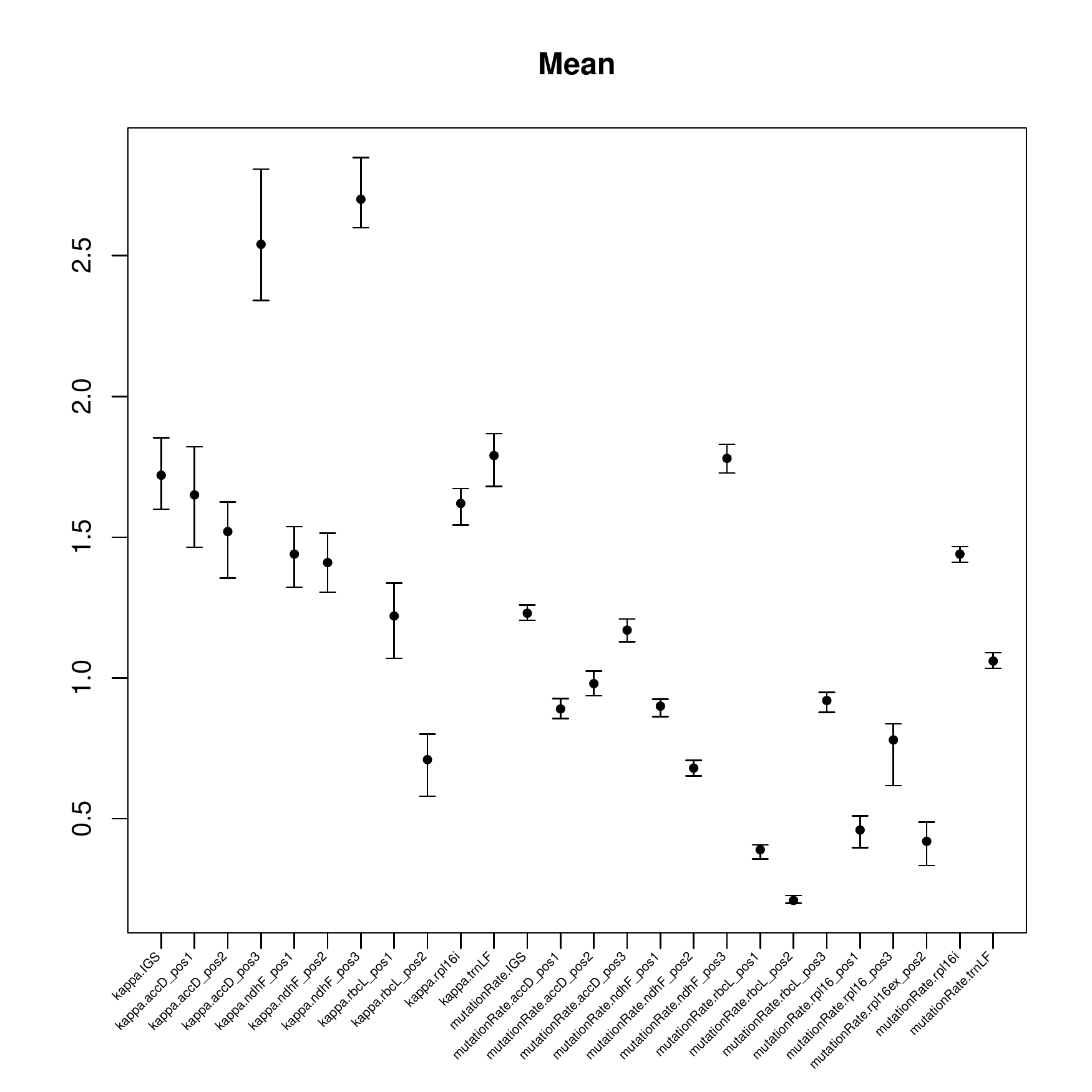}
	\includegraphics[scale=0.45,clip=true,angle=0]{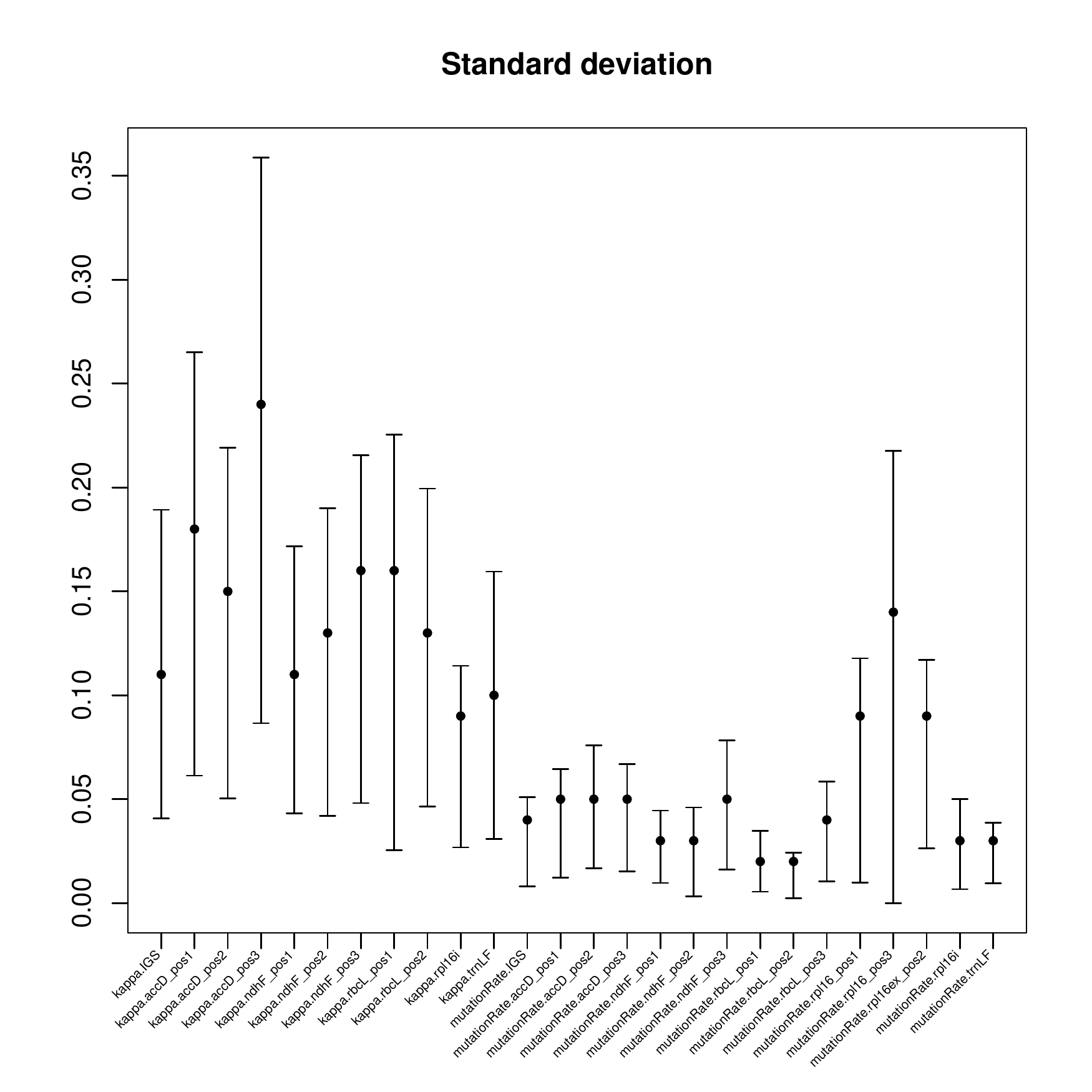}
	\caption{NS and MCMC estimates of the mean and standard deviation for some parameters in the Euphorbia dataset.  The intervals are inferred by using NS and the dots stand for the statistics obtained from the MCMC analysis.  All the NS intervals contain the MCMC statistics, suggesting its effectiveness in parameter inference.}
	\label{Figure2}
\end{figure}

The results are displayed in Table~\ref{Table2}. The NS posterior sample size fluctuated between 9 and 47 points, with a mean of around 15 and standard deviation of 7.3.  NS statistics also include their corresponding 95\% confidence intervals.  It is apparent that all the intervals contain the MCMC estimates for the means as well as for the standard deviations of the posterior distribution.  These include posterior, likelihood, and prior values as well as some parameters.  Figure~\ref{Figure2} shows the 95\% confidence intervals for the NS estimates for some other parameters which are in similar scales.  The points stand for the MCMC estimates, which are all within the NS intervals.

\begin{table}
	\centering
	\scalebox{0.8}{
		\begin{tabular}{lrrrrrrrr}
			\toprule
			& \multicolumn{2}{c}{MCMC} & \multicolumn{6}{c}{NS} \\
			\cmidrule{2-3} \cmidrule(lr){4-9}
			& \multicolumn{1}{c}{Mean} & \multicolumn{1}{c}{SD} & \multicolumn{1}{c}{Mean} & \multicolumn{1}{c}{lower} & \multicolumn{1}{c}{upper} & \multicolumn{1}{c}{SD} & \multicolumn{1}{c}{lower} & \multicolumn{1}{c}{upper} \\
			\midrule
			Posterior & -66831.62 & 17.99 & -66838.90 & -66904.90 & -66772.91 & 14.36 & 0 & 35.51 \\
			Likelihood & -67524.00 & 16.32 & -67531.22 & -67602.26 & -67460.19 & 13.57 & 0 & 37.48 \\
			Prior & 692.39 & 7.99 & 692.32 & 683.45 & 701.18 & 5.29 & 0.99 & 9.59 \\
			TreeHeight & 0.04 & 0.00 & 0.04 & 0.04 & 0.05 & 0.00 & 0.00 & 0.00 \\
			YuleModel & 728.35 & 7.89 & 728.33 & 719.28 & 737.37 & 5.39 & 1.04 & 9.74 \\
			birthRate & 112.63 & 9.20 & 113.22 & 105.35 & 121.09 & 7.06 & 1.34 & 12.77 \\
			kappa.rbcL\_pos3 & 5.04 & 0.45 & 5.09 & 4.75 & 5.43 & 0.36 & 0.14 & 0.59 \\
			kappa.rpl16\_pos1 & 1.34 & 0.47 & 1.47 & 1.07 & 1.87 & 0.42 & 0.16 & 0.67 \\
			kappa.rpl16\_pos3 & 3.85 & 1.21 & 4.15 & 3.07 & 5.23 & 1.04 & 0.16 & 1.93 \\
			kappa.rpl16ex\_pos2 & 7.79 & 3.67 & 7.33 & 4.52 & 10.14 & 2.55 & 0.16 & 4.95 \\
			ucldStdev & 0.57 & 0.03 & 0.56 & 0.52 & 0.60 & 0.02 & 0.01 & 0.04 \\
			rate.mean & 0.93 & 0.04 & 0.93 & 0.89 & 0.98 & 0.02 & 0.00 & 0.04 \\
			rate.variance & 0.38 & 0.06 & 0.38 & 0.32 & 0.43 & 0.04 & 0.01 & 0.07 \\
			rate.coefficientOfVariation & 0.62 & 0.04 & 0.61 & 0.57 & 0.65 & 0.03 & 0.01 & 0.05 \\
			\bottomrule
		\end{tabular}
	}
	\caption{MCMC and NS posterior means and standard deviations of estimated features for the Euphorbia dataset -- see Figure \ref{Figure2} for remaining kappa and relative substitution rate estimates.  NS estimates also include their corresponding 95\% confidence intervals, which in each case contain the MCMC statistics.}
	\label{Table2}
\end{table}

\subsection{Tetrapod}

The DS1 alignment from \citet{Hohna:2012} consists of 27 ribosomal RNA sequences of tetrapod with 1,949 nucleotides
\citep{Hedges:1990}.  Its most remarkable feature is its tree space, which contains separate regions which form ``islands" with high posterior probabilities \citep{Hohna:2012}.  \cite{Whidden:2015} showed that two of these islands are separated by only 2 SPR operations, but that the intermediate topology is so unlikely that it was never visited in their MCMC analysis.  In general, the MCMC chains tend to get stuck in one of these tree islands, \skadd{a common problem for standard MCMC methods.}

These characteristics make DS1 a good case to assess the performance of NS and compare it to standard methods.  We evaluate its attributes for parameter inference and marginal likelihood estimation.  It will be assumed a GTR and a relaxed clock models are suitable.  As prior for the tree topologies, a Yule model is assumed with a Uniform(0, 150) for its birth rate; for the clock rate a lognormal distribution; for the relative rates, the default Gamma priors implemented in BEAST 2.

\subsubsection{Parameter inference}

Two independent MCMC analyses are performed in order to sample from the posterior distribution.  Each chain has a length of 30 million with a burn-in period of 33\%.  The samples are taken every 10,000 steps.  Furthermore, two independent NS are performed in order to sample the posterior.  For each run, 100 active points are considered with 20,000 steps in order to generate the independent points required by NS at each iteration.

\begin{figure}
	\centering
	\includegraphics[scale=0.45,clip=true,angle=0]{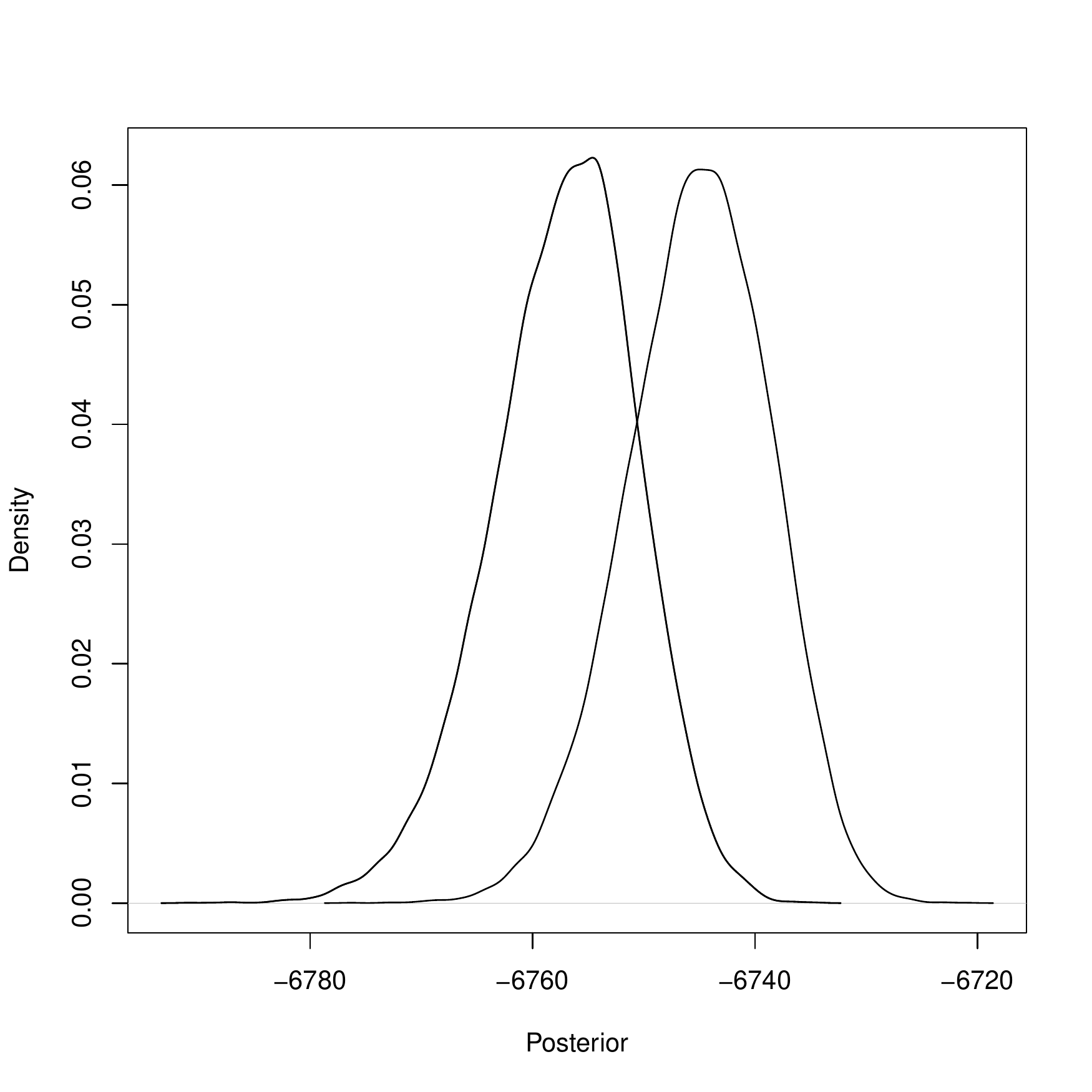}
	\caption{Posterior densities obtained from two independent MCMC chains for the Tetrapod dataset, converging to different distributions due to the tree islands in the parameter space.}
	\label{Figure3}
\end{figure}

\begin{figure}
	\centering
	\includegraphics[scale=0.45,clip=true,angle=0]{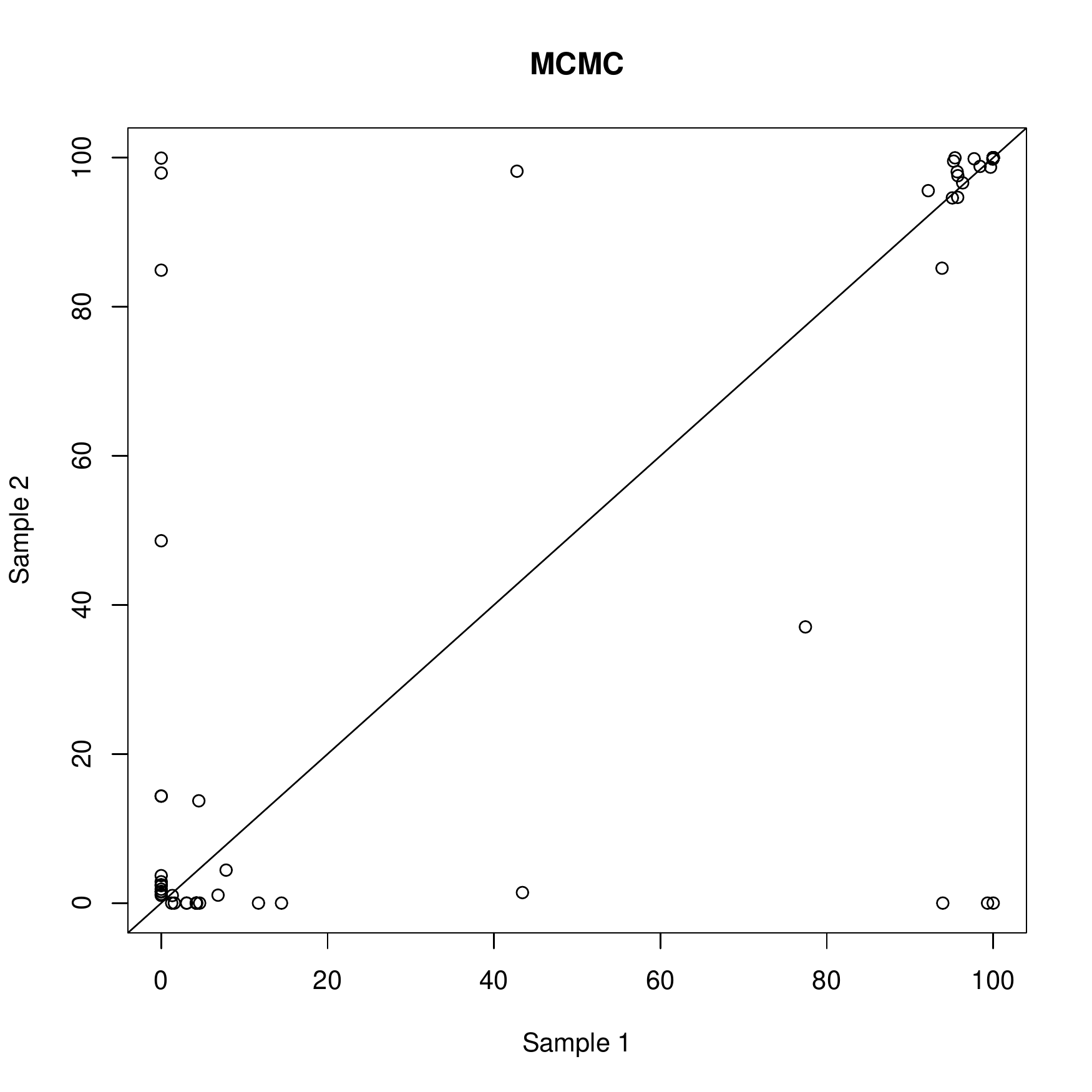}
	\includegraphics[scale=0.45,clip=true,angle=0]{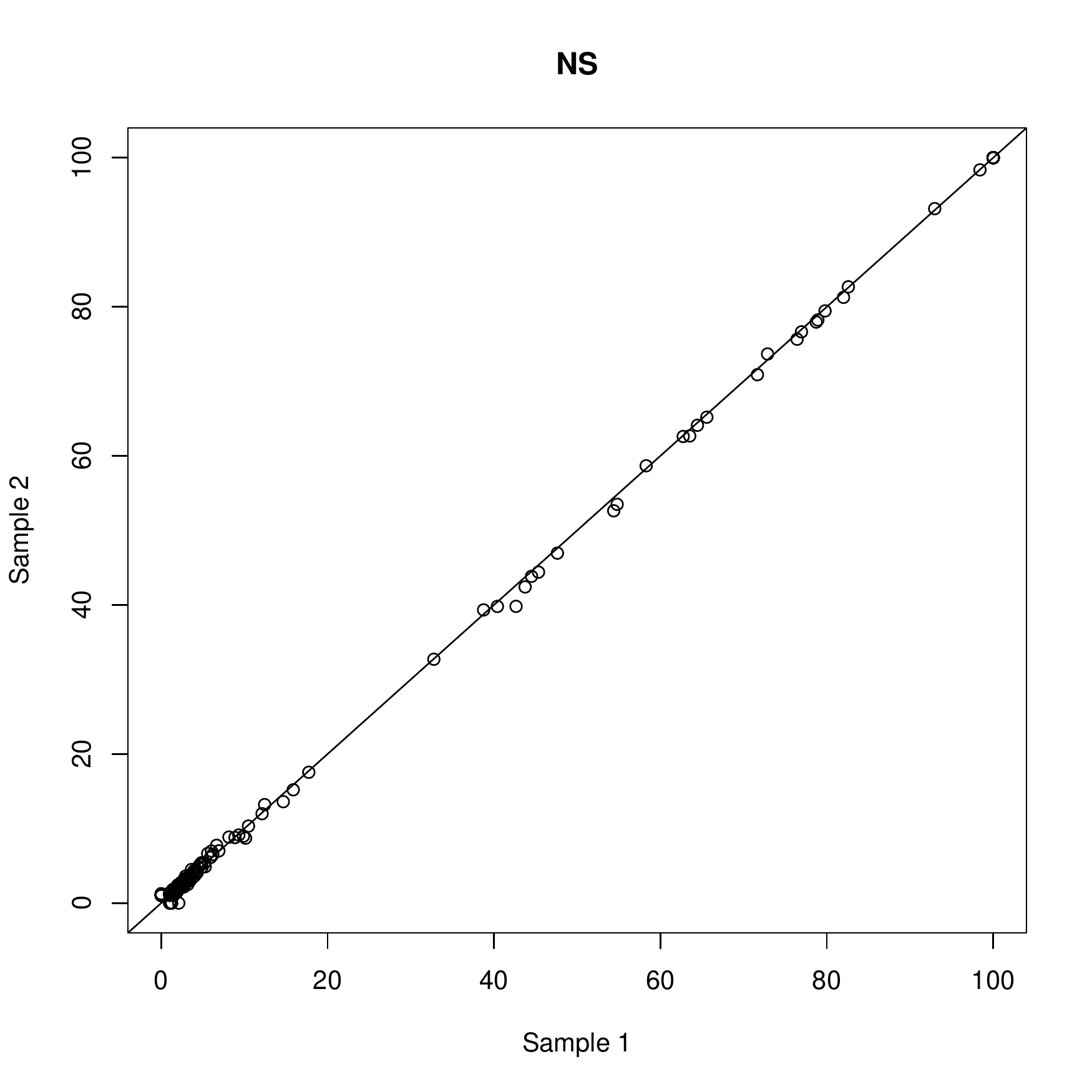}
	\caption{Comparison of 2 MCMC and NS posterior clade probabilities for the Tetrapod dataset. The graph on the left shows the difficulty of sampling a parameter space which contain tree islands, where the MCMC chains tend to get stuck and consequently converge to different distributions.  The graph on the right shows consistency of NS yielding posterior samples in independent runs, unlike standard MCMC methods.}
	\label{Figure4}
\end{figure}

The posterior values for the 2 MCMC chains are displayed in Figure~\ref{Figure3}.  It is apparent that they converge to different distributions. It is highly probable that the Markov chains got stuck on one of the tree islands, being unable to escape. Figure~\ref{Figure4} shows their posterior clade probabilities which reflect how different the samples are.  Note that there are clades with 100\% support in one sample, whereas in the other sample, the same clades have 0\% support.  On the other hand, NS yields samples with consistent clade probabilities (Figure~\ref{Figure4}).  The low values are notably in total agreement.  They tend to differ in case of problems in the sampling.  NS posterior sample sizes in this example were around 2,300.  The analysis was replicated multiple times, obtaining similar posterior clade probabilities from the NS analyses (not shown).

NS has the potential of exploring the parameter space in different areas at each iteration, granted by the active points.  Even in the case of using a single active point, the particle explores  the parameter space according to the prior, with the likelihood restriction, but not according to the posterior, which embodies the difficulties.


\subsubsection{Marginal likelihood estimation}

The marginal likelihood is estimated by means of SS and NS.  For the SS algorithm, the estimation is carried out by using 384 and 2,560 steps.  In general, these specifications should be enough to yield reliable estimates.  For NS, 1 and 100 active points are considered, respectively.

Table~\ref{Table3} displays the results.  For the two different specifications in SS, the estimates are quite similar which could lead one to trust the outcome.  However, these values are more than 2 SDs below that of NS.  Presumably, SS has some problems sampling near the posterior distribution as was observed in the MCMC analyses carried out before, which could lead to the underestimation.  This is also a situation in which GSS would fail in estimating the marginal likelihood, due to the incapacity of the MCMC methods in approximating the posterior and consequently, in generating the reference distribution (see \hyperref[ex:stat_example]{Appendix} for an example which illustrates this situation).

\begin{table}
	\centering
	\begin{tabular}{cccccc}
		\toprule
		Method & Steps  	& Samples	&  \multicolumn{3}{c}{$\log \widehat{\mathcal{Z}}$} \\
		\midrule
		SS & 384		& 500 		&-6956.49 & -6959.06 & -6955.41 \\
		SS & 2560	& 500 		&-6956.99 & -6958.36 & -6953.69 \\
		\bottomrule
		%
		\toprule
		& $N$ & $H$ & SD & Iterations & $\log \widehat{\mathcal{Z}}$ \\
		\midrule
		NS & 1 	&  149.36 & 12.22 & 183    & -6952.69 \\
		NS & 100 &  146.90 & 1.21  & 17,754 & -6950.93 \\
		\bottomrule
	\end{tabular}
	\caption{Marginal likelihood estimates obtained through SS and NS under different specifications for the Tetrapod dataset.  \pmr{For SS, the 3 values for each specification stand for independent estimates.}  SS does not yield estimates close to the NS ones \pmr{(in terms of SD)}, possibly due to its difficulty in sampling near the posterior distribution.}
	\label{Table3}
\end{table}


\subsection{Chloridoideae}

The DS9 alignment \citep{Hohna:2012}  consists of 67 sequences of Chloridoideae with 955 nucleotides \citep{Ingram:2004}.  Its tree space is rather flat which \skadd{requires a} big sample size in order to \skadd{correctly} infer the posterior probabilities \citep{Hohna:2012}.  In this context, a Markov chain of length 10,000,000, with a burn-in period of 10\% and thinning factor of 10,000, is compared to an NS posterior sample considering 100 active points.  In particular, we are interested in how well NS performs for low probability clades.

\begin{figure}
	\centering
	\includegraphics[scale=0.45,clip=true,angle=0]{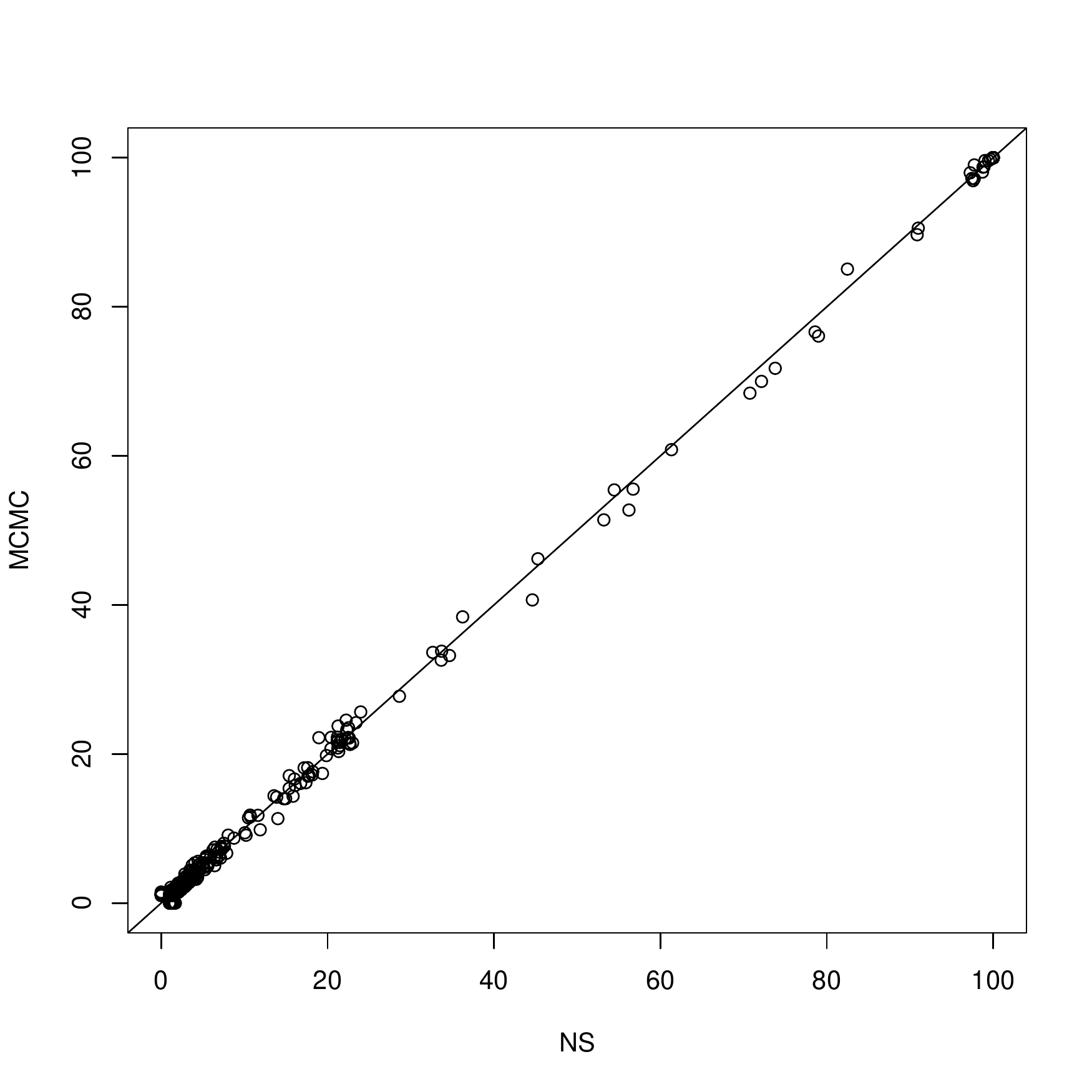}
	\caption{Posterior clade probabilities obtained via MCMC and NS for the Chloridoideae dataset.  Both samples are in agreement, showing the proficiency of NS in sampling the tree space.}
	\label{Figure5}
\end{figure}

The results are displayed in Figure~\ref{Figure5}.  The MCMC analysis has effective sampling sizes of at least 400, indicating a reliable approximation of the posterior.  The posterior clade probabilities obtained from NS are quite similar to those obtained from the MCMC analysis.  These values are located around the straight line, reflecting the agreement between the tree posterior samples.  Principally, the low probabilities are highly correlated, values which tend to differ in case of disagreement between the samples.  NS posterior sample sizes were of around 3,500.

\section{Conclusion}

Nested sampling is a general Bayesian algorithm that provides the means to estimate marginal likelihoods and to carry out parameter inference.  We have introduced it to \skadd{phylogenetic inference under variable tree topology}. Its performance has been compared to established methods available in many phylogenetic software packages.

NS performance has been assessed in different and challenging phylogenetic contexts. \pmr{Firstly, we verified that NS behaves well in a 4 taxa case, where it is possible to compare NS with the ILP method, and we found NS yields reliable estimates.}  Secondly, we compared NS to SS using a dataset which contain 197 \skadd{sequences} and 15 partitions.  In particular, we compared a strict clock model to a relaxed clock model, showing that NS with a single active point was enough to carry out model selection.  In this analysis, the relaxed clock model was found to fit the data better.  Then, we showed that the single NS run used to estimate the marginal likelihood for this model provides also the means to estimate the means and standard deviations of its posterior distribution.  Thirdly,  we showed that NS yields consistent posterior clade probabilities in independent runs in the case that the tree topology space contain tree islands, unlike standard MCMC methods.  In this context, we also showed that NS can differ from SS in marginal likelihood estimation, which can potentially have problems dealing in this scenario.  Finally, we evaluated NS performance in a flat tree parameter space, showing that NS posterior clade probabilities are in agreement with those obtained from a standard MCMC method.

PS and SS have become popular because of their high accuracy estimating the marginal likelihood.  Specifically, SS has become popular due to its implementation in widely-used phylogenetic software packages.  However, they rely on several problem-specific tuning parameters which need to be specified by the user, namely number and distribution of the $\beta$ values, number of samples from each transitional distribution and burn-in periods.  GSS dispenses with the distribution of the $\beta$ values, but it requires a number of posterior samples to calibrate the reference distribution.  These calibrations are essential to get good estimates.

\pmr{On the other hand, NS only requires the number of active points and the number of MCMC steps used to generate the replacement points.  The latter should be chosen in relation to the number of parameters.  This number should exceed the dimension of the parameter space in order to guarantee the generation of an independent point, otherwise the marginal likelihood estimate will be biased lower.  In our experience, it should be at least 10 times more than the number of parameters, but it is recommended to try different values.  The number of steps can be lower if the number of active points is increased.  Another area which remains open for further research is in finding more efficient transition kernels.  The length of the NS run is determined by the termination conditions described previously.  Overall, NS is in practice more user-friendly than the other methods presented.}


One of the most important attributes of NS is that it not only yields a marginal likelihood estimate but also provides a measure of its uncertainty in a single run.  \pmr{Even though the power posterior methods, at least SS and PS in their simple form, have an approximation for their uncertainties, these quantities are never reported in practice.}  
NS uncertainty is inversely proportional to the square root of the number of active points; the higher this number, the more accurate the estimate.  Under similar computational conditions, NS can have a higher uncertainty than GSS \citep{MaturanaPhDThesis}, though this depends on the problem and whether the GSS parameters are well-tuned.  However, its uncertainty can be calculated directly in a single run whereas GSS requires several replications to estimate its uncertainty, which can be carried out only after the reliability of the specifications for the tuning parameters have been tested.  In practice, one could use NS to estimate the marginal likelihood intervals for the competitive models, and thus carry out model selection based on them.  In the case that the intervals overlap, NS could be ran again for those specific models using more active points to increase the precision and narrow the intervals.

Nested sampling also provides the means to carry out parameter inference.  This does not involve an extra cost since the points used to estimate the marginal likelihood are recycled.  We have assessed the method to study clade probabilities and certain statistics of the posterior distribution.  In particular, we showed that NS, even in the case of using a single active point, can be used to generate confidence intervals for the parameters.  The method possesses interesting attributes.  For instance, unlike conventional MCMC methods, NS does not require a burn-in period.  In general, this period represents a high computational cost for MCMC methods.  Furthermore, the method explores the parameter space in a quite different way, which allows it to deal well in complex scenarios, such as those parameter spaces composed of tree islands, a challenging scenario for standard MCMC methods.

NS possesses several positive characteristics which make it a very competitive algorithm in comparison to the established methods used currently in phylogenetics.  It has been applied successfully to different fields and we believe this success can be replicated in phylogenetics as has been shown in this work.

The nested sampling algorithm is implemented in the NS package for BEAST 2, available from \url{https://github.com/BEAST2-Dev/nested-sampling} under the LGPL licence. Adapting an idea from \cite{MultiNest:2009}, a fully parallel version is implemented that runs $K$ nested sampling analyses with $N$ particles, but selects starting points from the pool of all available $K \times N$ active points (conditioned on having an appropriate likelihood to start with). The NS package allows for phylogenetic inference under any of the models available to BEAST 2.


\bibliographystyle{elsart-harv}
\bibliography{VNS_SystBiol}


\section*{Appendix}
\subsection*{Statistical example}
\label{ex:stat_example}

\cite{Skilling:2006} presented a statistical example in order to illustrate NS performance in the case that the likelihood (as a function of the cumulative prior probabilities) is partly convex.  This function was defined as the sum of two Gaussians centred at zero. The first has a standard deviation of 0.1 and the second a standard deviation of 0.01.  In addition, the second Gaussian has a factor of 100 which makes it contribute more to the shape of the
posterior distribution.  Thus, the likelihood is a relatively flat density with a spike in its centre.  This model, which includes a phase transition, poses problems to power posterior methods in their simple form, i.e., when the transitional densities define the path between the prior and the posterior.  On the other hand, they can perform well in their generalised forms, but if and only if an adequate reference distribution is used.

\cite{Maturana:2017} showed that a uniform works well as reference distribution in this example.  A tentative alternative would have been a normal distribution.  However, this poses some issues.  The standard MCMC methods used to draw from the posterior are highly dependent on the starting values.  If these are around zero, the MCMC chain will get trapped in the spike area, being almost impossible to escape from it.  As a result, the sample will fail in representing adequately the posterior and will lead to a poor reference distribution.  Actually, this will be centred at zero with a standard deviation of 0.01, as the narrow Gaussian in the likelihood function.  Thus, the reference distribution will only encapsulate the area where the spike is located and leave without consideration the rest of the parameter space, which will have a direct effect on the marginal likelihood estimation.  In their particular case, the excluded areas do not contain a significant volume and therefore the bias in the estimate would be small.

That analysis prompted the study of the same model, but where the excluded areas contain a bigger volume.  For this, we consider a different version of the statistical model analysed by \cite{Skilling:2006}.  In our model the likelihood is composed of the sum of the same two Gaussians, but with the factor which scales the spike reduced to 1, that is
\begin{align}
\label{eq:convex_like}
L(\bm{\theta}) = \prod_{i=1}^{d} \dfrac{1}{v \sqrt{2 \pi}} \exp \bigg( -
\dfrac{\theta_{i}^{2}}{2 v^{2}}\bigg) + \prod_{i=1}^{d}\dfrac{1}{u \sqrt{2
		\pi}} \exp \bigg( - \dfrac{(\theta_{i} - \mu)^{2}}{2 u^{2}}\bigg),
\end{align}
where $\bm{\theta}$ is a $d$-dimensional parameter vector, $d=20$, $\mu=0$, $v=0.1$ and $u=0.01$.  We consider a uniform prior in the unit cube $[-0.5, 0.5]^{d}$ for $\bm{\theta}$.  Thus, the marginal likelihood is 2.

\begin{figure}
	\centering
	\includegraphics[scale=0.46,clip=true,angle=0]{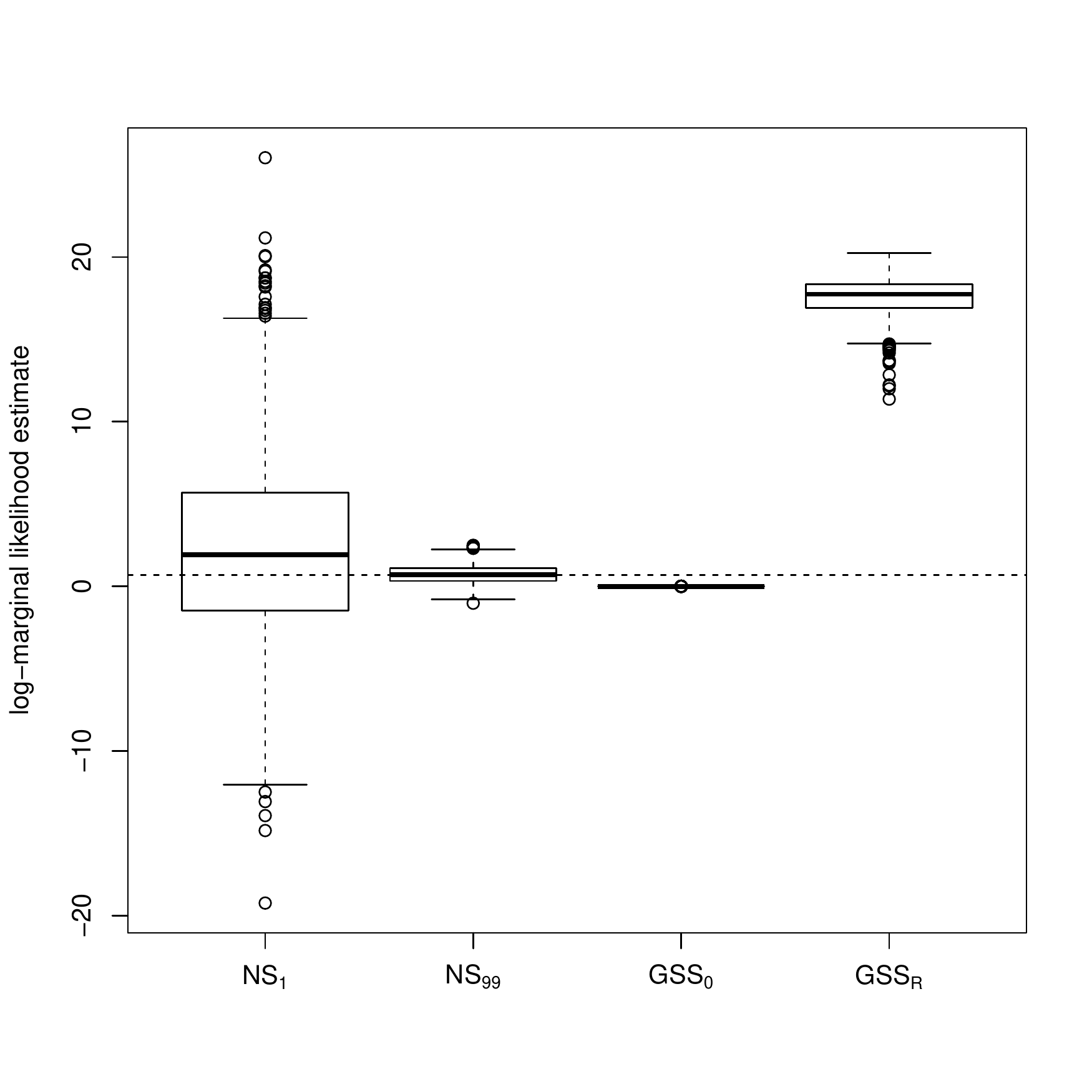}
	\includegraphics[scale=0.46,clip=true,angle=0]{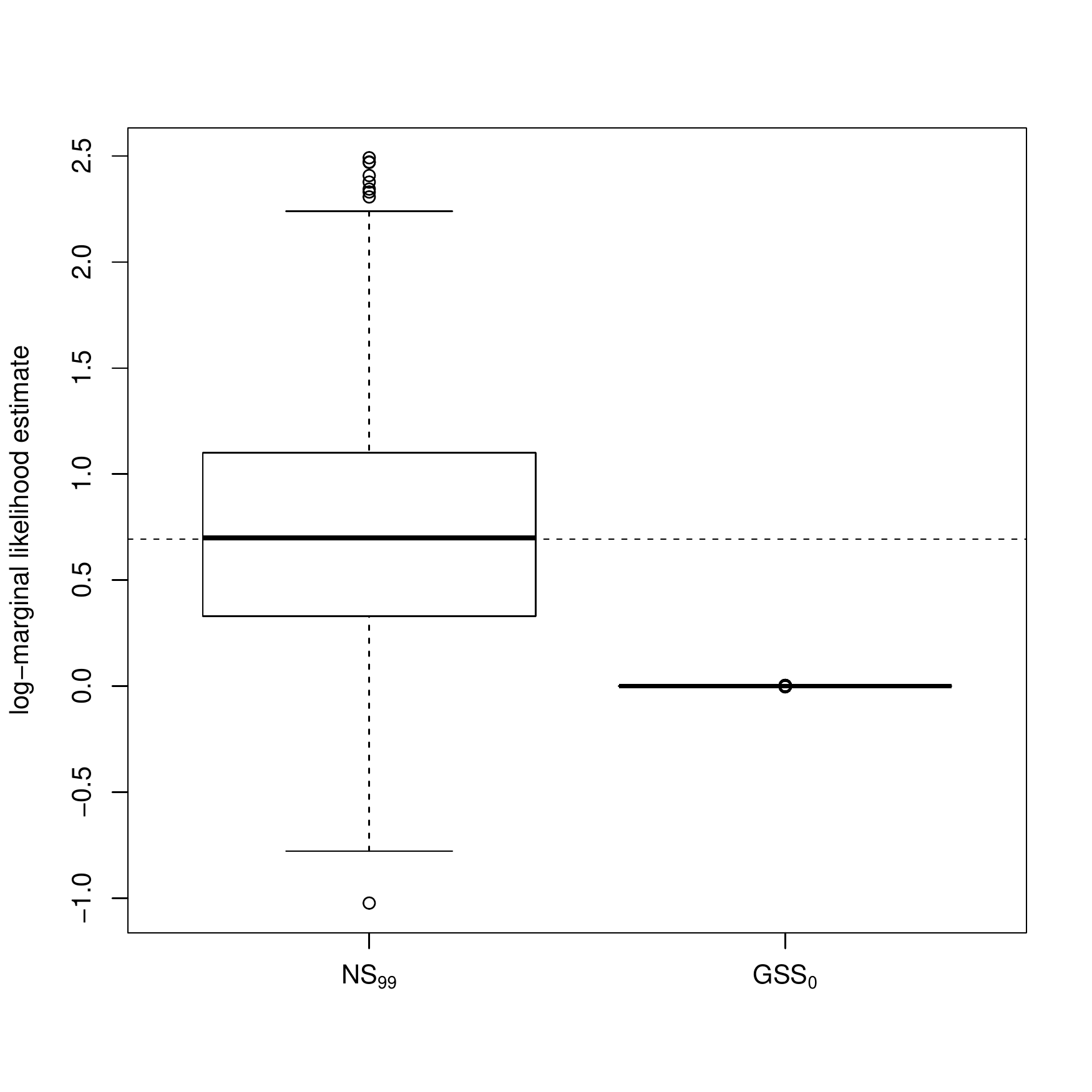}
	\caption{NS and GSS marginal likelihood estimates in the statistical example.  The subscripts in the NS methods stand for the number of active points, whereas in the GSS method depict the starting value specifications to generate the posterior samples (see text for more details).  The horizontal dotted line stands for the true log marginal likelihood value.}
\label{Figure6}
\end{figure}

We assess the marginal likelihood estimation by using NS and GSS.  For NS, we use 99 active points which makes it require around 10,000 iterations/samples.  We also include the NS estimate with only a single active point to evaluate its performance in the simplest condition.  To calibrate the reference distribution required by GSS, we consider two approaches when generating the posterior samples: i) starting the MCMC chain at zero and ii) starting at random values drawn from a Uniform(-0.5, 0.5).  The estimates in both cases will be referred as $\text{GSS}_0$ and $\text{GSS}_{\text{R}}$, respectively.  For both approaches, we use 1,000 posterior samples to parameterise the normal distribution, 100 transitional distributions and 100 samples from each of them.  This yields a total of 10,000 samples to calculate the GSS estimate.  This is without considering the initial posterior samples.  We use slice sampling to generate the samples.  The $\beta$ values are chosen according to evenly spaced quantiles of a Beta(0.3, 1.0) distribution, and following a ``melting" scheme, that is, starting from the posterior and moving down to the reference distribution.  The estimations are replicated 1,000 times and are displayed in Figure~\ref{Figure6} in where the horizontal dotted line stands for the the true value $\log(2)=0.693$.

\begin{figure}[]
	\centering
	\includegraphics[scale=0.46,clip=true,angle=0]{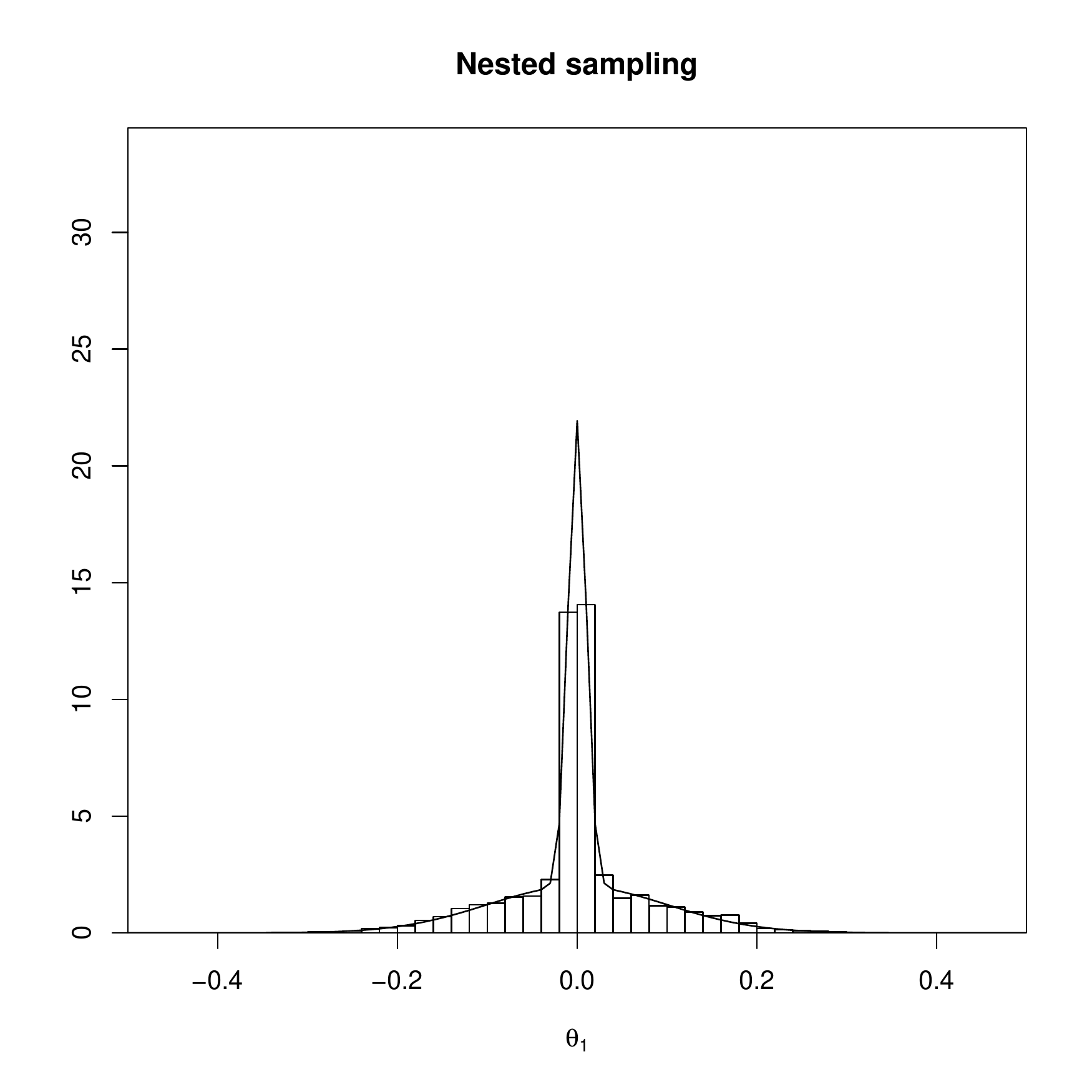}
	\includegraphics[scale=0.46,clip=true,angle=0]{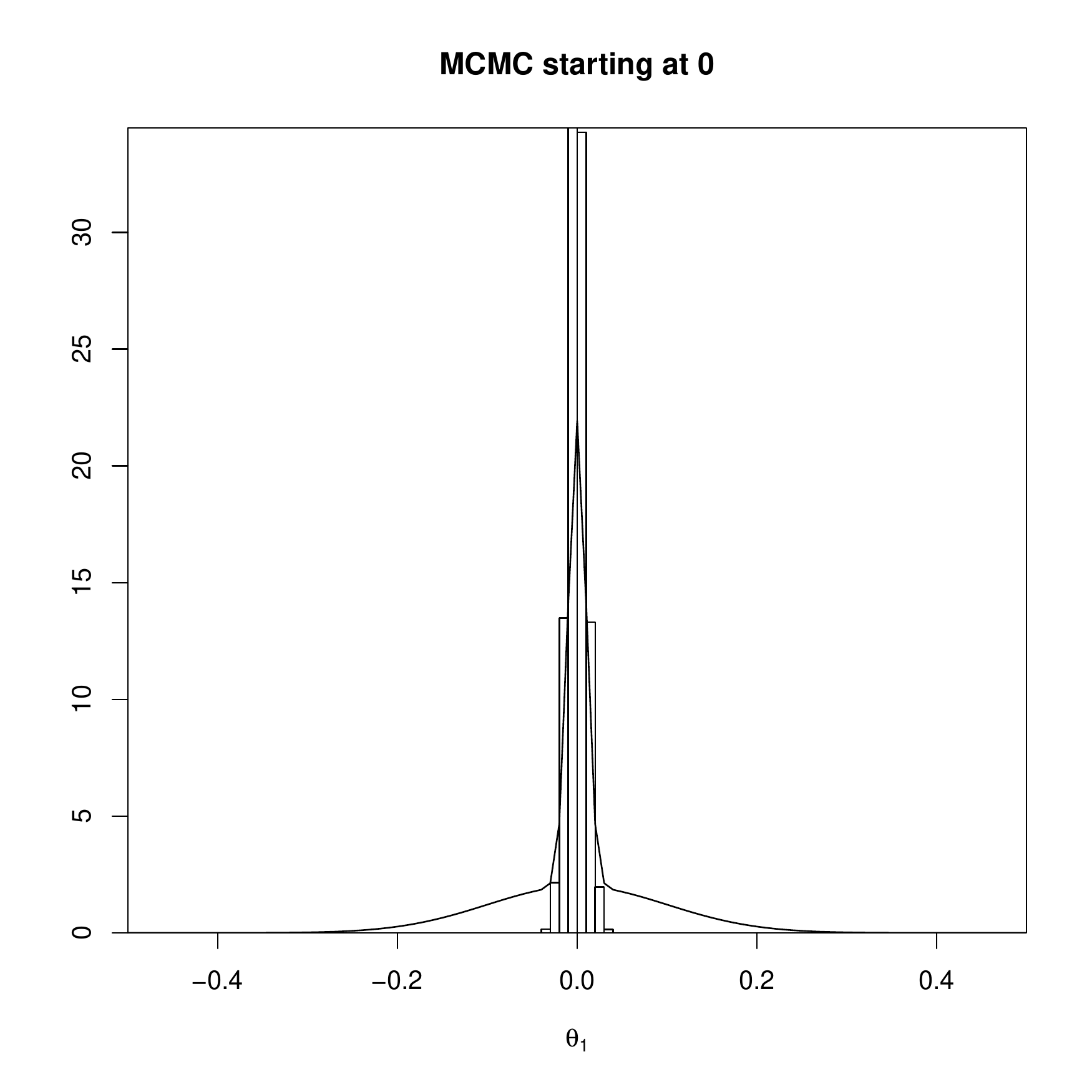}
	\includegraphics[scale=0.46,clip=true,angle=0]{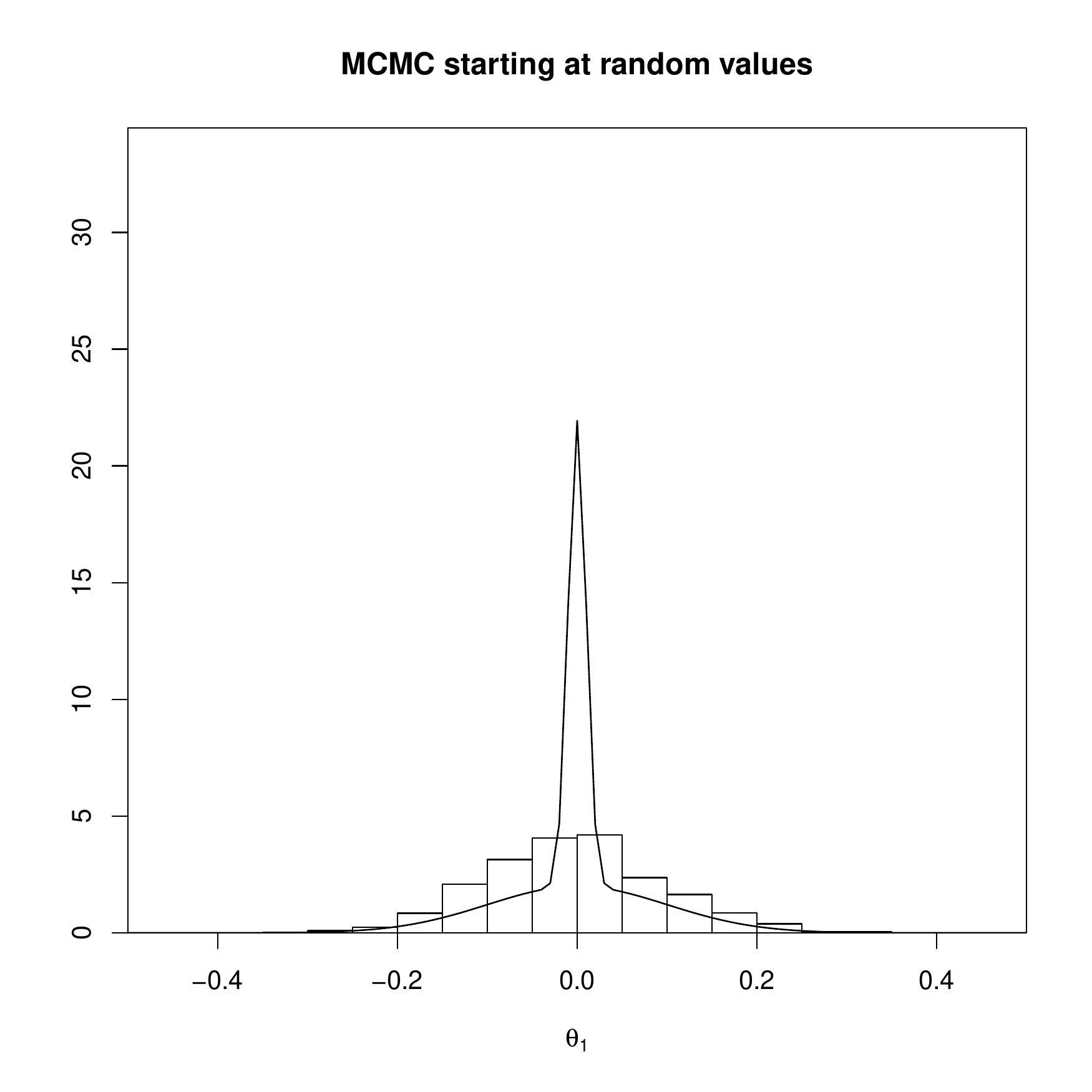}
	\caption{The histograms stand for 10,000 posterior samples for the first component of $\bm{\theta}$ by using nested sampling, and MCMC samples using two different starting values.  This behaviour is similar in all the components of $\bm{\theta}$.  The continuous lines depict the true marginal density.}
	\label{Figure7}
\end{figure}

$\text{GSS}_0$ underestimates slightly the true value.  Its estimates are around 0 with a standard deviation of 0.001.  Its reference distributions are centred approximately around~0 with a standard deviation of 0.01.  Consequently, they restrict their samples to the interval $[-0.03,0.03]$ excluding those areas which now, unlike in the original example, have a larger amount of probability mass.  This is clearly illustrated in one dimension posterior sample in Figure~\ref{Figure7}.  These significant areas are excluded from the marginal likelihood estimation.  This is the reason of the underestimation  which is now much more severe than in the original model.  Even in the case of increasing significantly the number of steps, the $\text{GSS}_{\text{0}}$ estimates do not change (results not shown), with which one would be tempted to trust in the reliability of the estimate.  On the other hand, $\text{GSS}_{\text{R}}$ overestimates the true value due to the failing of its reference distribution on pondering the different areas of the parameter space.  Its estimates have a mean of 17.56 and a standard deviation of 1.18.  Even in the case of increasing significantly the number of steps, the estimates are far away from the true value (results not shown).

On the contrary, NS estimates are around the true value.  Even in its simple case, with a single active point.   The estimates have a standard deviation of 5.85 and 0.57 for the case of 1 and 99 active points, respectively.  NS also provides accurate posterior samples which cannot be obtained by conventional MCMC methods (see Figure~\ref{Figure7}), such as Metropolis-Hastings or slice sampling.


Due to the likelihood shape, the approximation of the reference distribution for this model is determined by the starting point in the Markov chain  (see Figure~\ref{Figure7}).  If all the components of the starting point are around 0, the distribution will be approximately a N(0, 0.01) for each component of $\bm{\theta}$, but if there is at least one of them a bit far from its centre, let say outside the approximated interval $[-0.023, 0.023]$, the reference distribution for each component will be approximately a N(0, 0.1).  Actually, they are the plateau and the narrow normal distributions, respectively, which compose the likelihood.  The consequences of failing to approximate the posterior distribution directly impacts the marginal likelihood estimates.  Therefore, in situations where the posterior is not an easy distribution to sample from, (for instance, Tetrapod dataset), GSS should not be used.


\end{document}